\documentclass[aps,prx,reprint,twocolumn,superscriptaddress,longbibliography]{revtex4-2}
\usepackage{amsmath,amssymb,amsthm}
\usepackage{bbm}
\usepackage[ruled,vlined,linesnumbered]{algorithm2e}
\usepackage{physics}
\usepackage{graphicx}
\usepackage{xcolor}
\usepackage{natbib}
\usepackage[pdftex, colorlinks = true,hypertexnames=false]{hyperref}
\hypersetup{colorlinks,
    linkcolor=red,
    citecolor=blue,      
    urlcolor=brown,
}

\usepackage{comment}

\newtheorem{thm}{Theorem}
\newtheorem{lem}{Lemma}
\newtheorem{fact}{Fact}
\newtheorem{defn}{Definition}

\makeatletter
\def\thm@space@setup{\thm@preskip=1pt
\thm@postskip=1pt}
\makeatother

\newenvironment{sketch}{%
  \proof}{\endproof}

\DeclareMathOperator{\sign}{sgn}
  
\usepackage{times}
\normalfont
\usepackage[T1]{fontenc}

\usepackage{float}
\usepackage{mathpazo}

\begin{document}

\title{Efficient Quantum Voting with Information-Theoretic Security}

\author{Emil T. Khabiboulline}
\thanks{These authors contributed equally.}
\affiliation{Department of Physics, Harvard University, Cambridge, Massachusetts 02138, USA}
\author{Juspreet Singh Sandhu}
\thanks{These authors contributed equally.}
\affiliation{John A. Paulson School of Engineering and Applied Sciences, Harvard University, Cambridge, Massachusetts 02138, USA}
\author{Marco Ugo Gambetta}
\affiliation{QMATH, Department of Mathematical Sciences, University of Copenhagen, DK-2100 Copenhagen \O, Denmark}
\author{Mikhail D. Lukin}
\affiliation{Department of Physics, Harvard University, Cambridge, Massachusetts 02138, USA}
\author{Johannes Borregaard}
\affiliation{Department of Physics, Harvard University, Cambridge, Massachusetts 02138, USA}
\affiliation{QuTech and Kavli Institute of Nanoscience, Delft University of Technology, Lorentzweg 1, 2628 CJ Delft, The Netherlands}

\begin{abstract}
    Ensuring security and integrity of elections constitutes an important challenge with wide-ranging societal implications. Classically, security guarantees can be ensured based on computational complexity, which may be challenged by quantum computers. We show that the use of quantum networks can enable information-theoretic security for the desirable aspects of a distributed voting scheme in a resource-efficient manner. In our approach, ballot information is encoded in quantum states that enable an exponential reduction in communication complexity compared to classical communication. In addition, we provide an efficient and secure anonymous queuing protocol based on a distributed secure sum. As a result, our scheme only requires modest quantum memories with size scaling logarithmically with the number of voters. This intrinsic efficiency together with certain noise-robustness of our protocol paves the way for its physical implementation in realistic quantum networks.
\end{abstract}

\maketitle

\section{Introduction} \label{sec:Introduction}

 Voting protocols have been studied extensively in classical cryptography~\citep{bernhard2013cryptographic} and numerous classical voting protocols have been developed~\citep{fujioka1992practical, bernhard2011adapting, bernhard2012necessary, cortier2013attacking}. To have a fair and secure voting outcome, an ideal protocol should meet the following  requirements:
\begin{enumerate}
\item \emph{Correctness}: With no adversary interference, the protocol correctly counts all voters. \label{it:correctness} 
\item \emph{Accountability}: Only eligible voters are allowed to vote with only one vote per voter.
\item \emph{Verifiability}: Voters should be able to verify their vote and the outcome of the election.
\item \emph{Anonymity}: The identity of the voter cannot be compromised. \label{it:anonymity}
\end{enumerate}

\begin{figure}[ht!]
\centering
\includegraphics[width=0.45\textwidth]{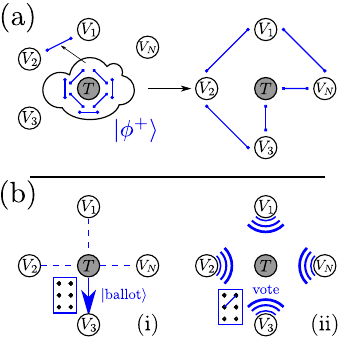}
\caption{Overview of the voting protocol. (a) A tallyman $T$ distributes entangled pairs $\ket{\phi^+}$ to voters $V_i$, connecting themselves in a ring. (b) The shared entanglement enables (i) anonymous transmission of a quantum state encoding a collection of random bits (depicted as nodes of a graph) and (ii) anonymous broadcasting of votes, which are masked by a single obtained parity of two random bits (depicted as an edge on the graph).}
\label{fig:protocol}
\end{figure}

At present, the security of advanced voting protocols relies on  assumptions about the computational complexity of certain mathematical problems~\citep{Diffie1976}, such as integer factorization~\citep{Rivest1978}. However, some of these problems can be potentially solved efficiently on future large-scale quantum computers~\citep{Shor1994}. At the same time,  quantum systems may allow the sharing of quantum information between voters that can, in principle, avoid computational assumptions in favor of information-theoretic security~\citep{Ueli1999}. Even if the encrypted information is stored, no future advances in computational power will enable its decoding. This observation has spurred active development of quantum voting protocols, see, e.g.,~\citep{Arapinis2021} for a recent review. However, existing approaches are either inefficient or do not satisfy all desirable security criteria. While early contributions~\citep{Hillery2006,Vaccaro2007} explored the use of  distributed entangled states to compute the voting function, their security is not rigorously established, especially in practical settings. For instance, quantum communication that is exponential in the number of voters is required to prevent double-voting~\citep{Hillery2006,Bonanome2011,Arapinis2021}. Recent work~\citep{Centrone2021} proposes using quantum anonymous broadcast for voting. We introduce ballot states enabling votes to contribute to the election's tally. Without ballot states, an adversary could remotely take the positions of absentee voters who do not participate in the election. Other advantages include less reliance on physical security and on a certain number of voters being present, since the ballot states can be made with a larger parameter than that number. In addition, the protocol does not have to abort due to double-voting, and thus is more robust. Furthermore, we introduce a quantum secure summation protocol to queue voters and avoid collisions. In contrast, commonly used~\cite{Unnikrishnan2019} classical collision detection~\citep{Broadbent2007} relies on pairwise authenticated channels, consuming more resources, and has not been proven secure against quantum adversaries~\citep{Lipinska2018}.

In this article, we describe a novel approach that exploits an exponential separation between quantum and classical communication complexity to authenticate voters and prevent forgery. Our scheme (\autoref{fig:protocol}) features a (generally untrusted) tallyman, who distributes ballots encoded in quantum states to the voters. We make use of a special encoding that has been shown to provide an exponential reduction in communication complexity~\citep{Yao1979} compared to classical encodings for solving a specific matching problem~\citep{Gavinsky2007}. This feature is related to a threshold~\citep{Shi2015} for locally decodable codes~\citep{Kerenidis2004} and serves as the basis for unforgeable quantum money~\citep{Gavinsky2012}. We use the communication advantage to ensure efficient scaling of resources for a large number of voters. Specifically, the total number of qubits communicated in our scheme scales polynomially with the number of voters and the size of local qubit memories scales only logarithmically with the number of voters. Importantly, we prove that this specific encoding also prevents forgery of votes. More precisely, we show that the voter can only learn enough information to output one vote, via projective measurement that collapses the quantum state encoding ballot information.

To send information to and from voters anonymously, we develop a protocol to order participants in a queue such that their position is anonymous. The ballot state is distributed to voters in the order of the queue via quantum teleportation using preshared entanglement. In the case of an untrusted tallyman, anonymous state teleportation schemes~\citep{Christandl2005,Unnikrishnan2019} can be employed to ensure voter anonymity. Distributed computation of the voting function is enabled by anonymous broadcast of the voters' ballots and encrypted votes, also in the order of the queue. We generalize a previous single-bit scheme~\citep{Christandl2005} to allow computation of a distributed sum efficiently and prove its information security. It is used to queue the voters in advance by securely detecting collisions, ensuring the correctness of the anonymous state transfer and broadcast, as well as improving their efficiency. Note that compared to prior quantum secure summation protocols~\citep{Shi2016}, our approach shares the output with all parties while retaining efficiency. We develop and incorporate generalized testing
strategies for high-dimensional entangled states, which are used in our algorithms. In addition, the construction of an anonymous queue may be of interest in other protocols.

In our approach, the tallyman is required to create and teleport the ballot state to each voter using photonic channels. The voters should be able to do permutations of the computational basis states allowing them to measure the ballot state in different bases. We outline how this can be achieved using one quantum node per voter containing a number of memory qubits that only scales logarithmically with the total number of voters. Consequently,  quantum nodes with only tens of qubits are required for thousands of voters. 

The assumption in our scheme is an honest majority of tallymen. It is necessary in general to achieve multiparty secure computation~\citep{BenOr2006}. Complementary to that assumption, which can be achieved with an honest majority~\citep{BenOr2006}, is access to a trusted central source of randomness, which is necessary for efficient entanglement verification. For this purpose, we also assume pairwise private communication channels (a standard assumption~\cite{Broadbent2007}) between the tallymen and voters, which can be established with quantum key distribution. 
In the description of our protocol, for simplicity we describe perfect quantum operations. However, the security of our scheme is not broken in the presence of imperfections. Specifically, the degree of anonymity is reduced, which is quantifiable and known to the voters. Since real quantum networks generally have imperfect entanglement, this guarantee is important for implementation. In practice, given faulty operations, the scheme may need to be run more times to output the outcome without aborting.

Our main result can be summarized as (formalized later in \autoref{thm:main}): \newline \newline
    \emph{The desirable cryptographic properties (\autoref{it:correctness}~--\ \autoref{it:anonymity}) of a voting scheme can be satisfied for $N$ voters with information-theoretic security with efficient scaling involving total quantum communication of $O(N^5 \log N)$ qubits and $O(\log N)$ memory qubits per voter.}

\section{The Voting Scheme}

At a high level, the protocol consists of the following phases; for details, see the full algorithms listed in the supplement~\cite{supplement}:
\begin{enumerate}
    \item Ordering the voters anonymously in a queue.
    \item Distribution of ballots to voters in order.
    \item Writing up to one vote per ballot by the voters.
    \item Collection of ballots from voters in order.
    \item Verifying that voting ran correctly and aborting otherwise.
    \item Counting of votes.
\end{enumerate}
The first five steps are implemented with quantum algorithms. The last one is done classically. We first describe the specific steps of the voting scheme assuming a trusted tallyman. For readibility, the formation of the queue is described at the end. We then discuss the amendments to the scheme that allow for untrusted tallymen. Finally, we describe how the ballots need to be distributed anonymously. Concluding the section, we summarize the resource cost of the voting scheme.\newline 

\emph{Creating the quantum ballots.} The tallyman first samples a random bit string $x\overset{R}{\leftarrow}\{0,1\}^n$ of $n$ bits. As discussed later, $n=O(N^2)$, where $N$ is the number of voters, to ensure that the protocol runs correctly. For instance, $n=6$ and the sampled bit string is $x=101100$.

The tallyman then prepares $N$ copies of the ballot state
\begin{equation} \label{eq:ballot}
\ket{\psi_x}=\frac{1}{\sqrt{n}}\sum_{i=0}^{n-1}(-1)^{x_i}\ket{i}.
\end{equation}
Here $x_i$ is the $i$th bit of $x$. The quantum state can be encoded across $\lceil\log n \rceil$ qubits using a binary encoding of $i$. Following our example of $x=101100$, the state $\ket{\psi_x}$ is
\begin{equation} \label{eq:ballot2}
    \frac{1}{\sqrt{6}} (-\ket{000} + \ket{001} - \ket{010} - \ket{011} + \ket{100} + \ket{101}) \,,
\end{equation}
where $\ket{000}\equiv \ket{0}\otimes \ket{0} \otimes \ket{0}$ denotes a three-qubit state.

After creating the ballot states, the tallyman distributes one ballot state to each voter. Using quantum teleportation, only $\lceil\log n \rceil$ Bell pairs need to be shared between each voter and the tallyman due to the efficient qubit encoding of $x$.    
\newline

\begin{figure}[ht!]
\centering
\includegraphics[width=0.45\textwidth]{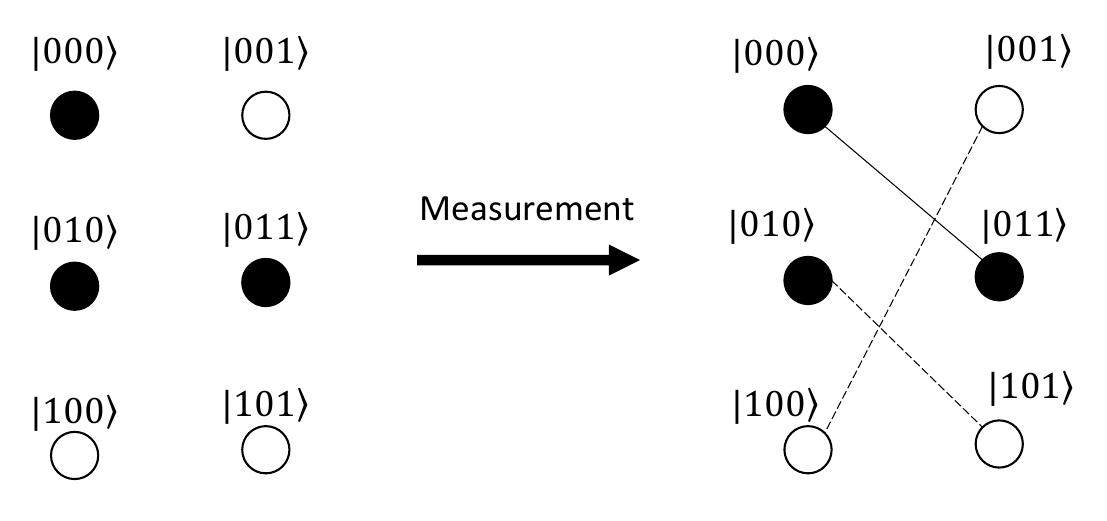}
\caption{Graph representation of the ballot state. \emph{Left:} The basis states are viewed as nodes that are either empty or filled corresponding to whether they encode a $0$ or $1$ bit of the bit string $x$. The depicted nodes correspond to the state in Eq.~(\ref{eq:ballot2}). \emph{Right:} The measurement performed by a voter corresponds to picking a random pairwise matching of the nodes and projecting onto one of the edges at random. The graph depicted corresponds to obtaining measurement outcome $\ket{000}+\ket{011}$ (solid line), having measured in the basis defined in Eq.~(\ref{eq:measbasis}).}
\label{fig:voting}
\end{figure}

\emph{Casting a vote.} Having received a quantum ballot state, a voter can, by means of a projective measurement, learn the parity of two random bits from the encoded bit string $x$. To illustrate this idea, we can view each of the basis states $\ket{i}$ as nodes. Measuring in the basis $\{\ket{i}\}$ would correspond to projecting at random on one of the nodes. Instead, a voter randomly chooses another complete measurement basis consisting of projectors on superpositions of two states $\frac{1}{\sqrt{2}}(\ket{i}\pm\ket{j})$ (where $i\neq j$). This can be viewed as picking a pairwise matching of all the nodes and projecting on a single random edge of the corresponding matching (see \autoref{fig:voting}). Following the above example, a measurement basis could, e.g., consist of projectors on states
\begin{eqnarray} \label{eq:measbasis}
\frac{1}{\sqrt{2}}\{\ket{000}\pm\ket{011},\ket{001}\pm\ket{100},\ket{010}\pm\ket{101}\} \,.   
\end{eqnarray}
By this measurement, the voter learns the parity between bit $x_i$ and $x_j$ since obtaining measurement outcome $\ket{i}\pm\ket{j}$ corresponds to the parity of $x_i$ and $x_j$ being even ($+$) or odd ($-$).

The voter classically encodes their vote $v\in\{0,1\}$ by calculating the vote bit $a=p \oplus v$, where $\oplus$ denotes addition modulo $2$. Here, we use the notion that even (odd) parity is assigned bit value $p=0$ ($p=1$). To compute the voting function, the voters broadcast their measured edge $(i,j)$ and their vote bit $a$ as described below. Once all voters have broadcasted their masked votes, the bit string $x$ will be revealed by the tallyman. This will enable the voters to decode each other's votes having the complete information of $x$. 

Importantly, $x$ is not to be revealed until after all voters have cast their votes (the case of a dishonest tallyman is treated below). In this way, a correct encoding and subsequent decoding of a vote can only be guaranteed if the voter can sample the correct parity of an edge from measurement of the ballot state. In what follows, we prove that given $k$ copies of the ballot state, a voter will only be able to obtain the parity of $k$ distinct edges (meaning that the edges do not share a node) in a deterministic fashion. Obtaining more than $k$ parities can only be achieved by randomly guessing a parity with a success probability of $1/2$. 

We use this property to rule out double voting by repeating the voting protocol $O(N)$ times and averaging the votes over the rounds. If any collection of $k$ voters (having access to $k$ ballot states) tries to output more than $k$ votes, they can only do so by randomly guessing parities for the extra votes, which will average to an even number of $0$ and $1$ votes over the repetitions and will cancel upon calculating the margin of the election.

The above arguments necessitate that only votes from distinct edges are counted. Otherwise, a vote can be forged, since $x_i \oplus x_j=(x_i \oplus x_k)\oplus(x_k \oplus x_j)$.  In the case that two voters share a node, only the first one broadcasted will be counted. The choice not to abort the protocol if overlapping edges appear prevents denial-of-service attacks, as detailed below, but poses the problem that two honest voters might sample overlapping edges by chance, which would result in one of the votes not being counted correctly. However, by increasing the length of the bit string $n$, the probability of this event can be made arbitrarily small~\cite{supplement}. Importantly, the number of qubits only scales as $\log n$, implying efficient use of quantum resources. 
\newline

\emph{Obtaining the election result.} The voters broadcast their encrypted votes and associated edge with multiple applications of a one-bit anonymous broadcast scheme~\citep{Christandl2005}; see details below. Each voter now has every other voter's information, but cannot decrypt it until the tallyman reveals $x$. With the tallyman's announcement, each voter classically computes parities and adds them to corresponding agreement bits:
\begin{eqnarray}
    p_i \oplus a_i &=& p_i \oplus (p_i \oplus v_i) \\
    &=& v_i \,.
\end{eqnarray}
The resulting votes are compiled in a function to decide the outcome of the election~\citep{supplement}. In order to ensure accountability (prevent double-voting), the election is repeated $O(N)$ times, averaging the tallies. Each voter then calculates the margin, which suffices to cancel any adversarial votes, and determines the winner in majority vote.  \newline

\emph{Anonymous queuing.} To retain privacy, the broadcast of the vote bits should be anonymous. For this, the one-bit anonymous broadcast scheme of Ref.~\citep{Christandl2005} is used repeatedly to broadcast both the edge and the agreement of a voter. The protocol assumes that the voters share a GHZ (Greenberger-Horne-Zeilinger) state. A GHZ state can be distilled from a ring of Bell pairs connecting the voters and the tallyman (see \autoref{fig:protocol}): performing a CNOT gate followed by a measurement on the target qubit combines Bell pairs into the nodes of a GHZ state, up to local single-qubit rotations~\citep{Komar2016}. Each GHZ state can be used to anonymously broadcast one bit of classical information. The anonymity is linked with the fidelity of the GHZ state (see \autoref{thm:anonymity}), which can be estimated by the voters through random checks of the correlations in the state~\cite{Pappa2012,Unnikrishnan2019,supplement}. In other words, the voters either choose to use a GHZ state for transfer or for testing. In this way, they can detect if an adversary tries to compromise anonymity, since the attempt will corrupt the correlations in the GHZ state. 

One complication of the above protocol is that one ring of Bell pairs only allows a single voter to broadcast one bit. However, to cast a vote, the voters need to broadcast both the edge $(i,j)$ of their ballot and their vote bit $a$, corresponding to $2\log n +1$ bits of classical information. Voters need to be queued such that the order in which the bits are broadcasted is known. At the same time, collisions where two voters try to broadcast simultaneously are prevented. Here, a position in the queue corresponds to a collection of 
Bell pairs used to broadcast the vote information of one voter. 

We solve this problem by introducing an anonymous queuing algorithm. It uses a multiparty secure sum that can be viewed as a generalization of the one-bit protocol of Ref.~\citep{Christandl2005}. Let us assume that the voters share entangled states of the form 
\begin{equation} \label{eq:queue}
\frac{1}{\sqrt{M}}[\ket{0}^{\otimes N}+\ldots+\ket{M-1}^{\otimes N}] \,. 
\end{equation}
Here, $M=2^{\lceil\log N\rceil}$ and the above state therefore corresponds to sharing $\lceil \log N \rceil$ GHZ states between the voters. The voter broadcasting applies $Z^a$, where $Z$ is the $M$-dimensional phase shift and $a\in\{0,\ldots,d-1\}$ is the announcement. Then, every voter applies the $M$-dimensional quantum Fourier transform on their share of the state, followed by measurement in the computational basis. If everyone announces their measurement outcome $k_l$, anyone can sum them together to obtain the announcement~\cite{supplement}:
\begin{equation}
    a=-\sum_{l=1}^N k_l \,.
\end{equation}
Thus, the correlations in the GHZ state act like a key. Explicitly, the voters could choose not to apply any phase flips and then the result would sum to $0$; individual terms would be the voters' share of the key, to which they could classically add their input. In either case, for multiple voters inputting an announcement, a sum is computed anonymously, up to a value $M$.

\begin{figure}[ht!]
\centering
\includegraphics[width=0.45\textwidth]{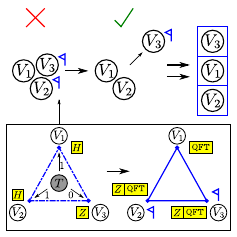}
\caption{Anonymous queuing. Voters attempt to enter the queue with the same probability, and input whether they attempt into an anonymous secure sum. If the sum evaluates to $1$, corresponding to an entry with no collisions, the respective voter enters the next open spot in the queue. The procedure continues with the remaining voters until everyone is queued. At the end, each voter knows only their own spot in the queue. Inset: To compute the distributed sum securely, the voters first verify that the shared state has high fidelity with an ideal GHZ state; then, they encode their input, masking with the verified correlations.}
\label{fig:queuing}
\end{figure}

The secure sum enables collision detection in the queuing algorithm. Namely, voters attempt to enter the queue with equal probability so that on average, there is no collision over the entire queueing process. Beforehand, each voter picks a random integer out of $d=O(N^2)$ possibilities, such that the protocol succeeds with constant probability~\citep{mitzenmacher2017probability}. The construction of the queue proceeds over $d$ rounds. To enter the queue in a certain round equal to their randomly chosen integer, a voter broadcasts $1$ instead of $0$. If the sum is greater than $1$, the queuing fails because several entries collided. Meanwhile, a sum of $1$ is viewed as a success (see \autoref{fig:queuing}). An outcome of $0$ results in an empty slot in the queue; at the end of the algorithm, these null entries are removed by the voters.

The anonymity of the queue is key to the security of the scheme. Namely, if the positions of voters are revealed, then an adversary learns which voter broadcasted a particular vote bit and edge. At the end of the election, they can check the voter's vote, similar to the honest voter. \newline

\emph{Generalized GHZ state verification.} In order to ensure anonymity, it is critical to verify that the shared states used by the voters are close to an ideal GHZ state (\autoref{eq:queue}). This requires running a verification protocol where each voter chooses a measurement based on the realization of a uniformly random bit string $\{X_i\}_{i=1}^N$ chosen by the tallyman \cite{Pappa2012}. If $X_i$ is $1$, voter $V_i$ performs a Hadamard operation; otherwise, the voter applies a Pauli $Z$ transformation. Then, measurements $Y_i$ are made in the $Z$-basis, and accepted \emph{if and only if},
\begin{equation}
    \sum_{i=1}^n Y_i = \frac{1}{2}\sum_{i=1}^n X_i \bmod 2\, .
\end{equation}
The above procedure is repeated for each of the $O(\log N)$ qubits making up the high-dimensional GHZ state, and the tallyman accepts if the test passes each time. The entire test is repeated $O(N)$ times to ensure desirable success probability.
\newline

\emph{Voting verification.} Anytime that voters notice deviation from the honest behavior of the protocol, they complain. The first trigger for a complaint is when GHZ state verification fails. The voters publicly announce the failure of the test and the GHZ state is not used. Second, at the end of the voting round, voters verify that their own vote is counted correctly and not flipped due to, e.g., a corruption of their ballot state. Here, anonymous complaints are filed by encoding into a secure sum, similar to what is described above, so the number of complaints per round is known. If, on average, the margin of the election is greater than twice the number of complaints, then the compromised votes would not affect the outcome and the election result holds. Otherwise, the election is aborted. \newline

\emph{Dishonest tallyman.} The tallyman has complete knowledge of $x$, so they have additional power compared to the voters. Specifically, a dishonest tallyman can output extra votes and might tamper with the states to compromise the anonymity of the voters. To avoid putting trust in the tallyman to behave honestly, we introduce amendments to the voting protocol, decentralizing the role of the tallyman. 

First, to circumvent extra votes outputted by the tallyman, the election is run in parallel with different tallymen with independent, private randomness. Each of these elections should yield the same result if the tallymen are honest, and we only need to assume the presence of an honest majority, deciding the outcome via majority vote. Any dishonest tallymen will be identified through the discrepancy in the outcome of their elections and there is thus a risk of exposure if a tallyman chooses to behave dishonestly. The independence in the randomness implies that dishonest tallymen cannot initiate attacks based on correlated randomness with honest tallymen.

Second, to ensure the anonymity of the voters, the ballot states are teleported anonymously to them. This way, the tallyman might tamper with a ballot state but will have no information about which voter received the tampered ballot. Consequently, anonymity will not be compromised. For the GHZ verification, assuming an honest majority of tallymen, we let one tallyman distribute the state and another pick the random bit string for testing if a voter wants to test, as described in the supplement~\cite{supplement}.   \newline

\emph{Anonymous Transfer.} To anonymously teleport the ballot states, we make use of anonymous quantum state transfer~\citep{Unnikrishnan2019}. Similar to anonymous broadcast, this protocol relies on shared GHZ states among the voters and the tallyman. From the GHZ state, a Bell pair is distilled between the tallyman and an anonymous voter, through which a qubit can be transferred by means of quantum teleportation. To distill a Bell pair, all but the tallyman and the receiving voter measure their qubits in the $X$ basis and broadcast the result (the receiver broadcasts a dummy bit to stay anonymous). As before, voters randomly run a verification protocol to ensure that the GHZ state is of high fidelity, which ensures anonymity.

One GHZ state enables only a single voter to receive one qubit anonymously. The anonymous queuing of the voters (as described above) allows the voters to know in which position 
they are to receive a ballot state and to avoid collisions. \newline

\emph{Resource requirement.} As detailed above, all subroutines of the voting protocol can be realized using shared Bell pairs between pairs of voters and the tallyman in a ring structure. The total resource consumption of one round (out of $O(N)$ rounds) of the voting protocol is $O(N^4\log N)$ qubits of communication~\citep{supplement}.

Importantly, the size of the local qubit memories of the voters and the tallyman only scales as $O(\log N)$ assuming a sequential operation of the voting scheme where Bell pairs are distributed at the start of each subroutine (e.g., ballot distribution, anonymous queuing) instead of in advance. In this way, the size of the memories will be determined by the ballot state and the anonymous queuing algorithm, which both require $O(\log N)$ memory qubits.
\newline

In the rest of this paper, we outline the elements of a physical implementation and describe the cryptographic properties of the scheme. The precise statements of the algorithms and technical material completing the proofs are relegated to the supplemental material~\citep{supplement}.

\section{Physical Implementation}

The execution of our voting scheme requires an underlying quantum network capable of distributing high-fidelity Bell pairs between adjacent voters in a ring structure. In summary, the scheme consists of the steps:
\begin{enumerate}
    \item First, anonymous queuing is run to establish the order of ballot distribution and vote casting. The tallyman repeatedly distributes $O(\log N)$ Bell pairs to the voters until the queuing is finished.  
    \item After the queuing, the tallyman again distributes $O(\log N)$ Bell pairs for anonymous state teleportation. Once an anonymous channel of $O(\log N)$ Bell pairs between the tallyman and a voter has been established, the tallyman creates a ballot state and teleports it to the voter.   
    \item Having received the ballot state, the voter measures in a random basis to calculate a (classical) bit used for encrypting votes.  
    \item Steps 2 and 3 are repeated until all voters have their vote bits. The tallyman now repeatedly distributes $O(\log N)$ Bell pairs to allow the voters to anonymously broadcast their encrypted votes one-by-one following the order dictated by the queue. Finally, the tallyman classically broadcasts the bit string $x$ used for decryption.     
\end{enumerate}
Note that the number of Bell pairs that a voter needs to simultaneously store and operate on scales as $O(\log N)$ due to the efficient encoding of the quantum ballot. Only modestly-sized local quantum processors are thus required even for a large number of voters. \newline

\emph{Entangled states.} The distribution of high-fidelity Bell pairs is a key functionality of quantum networks. In this application of voting, they are combined into GHZ states. Implementation with nodes based on atomic qubits allows for generation of entanglement that is deterministic, as has been experimentally demonstrated~\cite{Pompili2021}. Then, the success probability does not decay with the number of voters. Furthermore, the number of necessary Bell pairs to obtain a set target GHZ fidelity was found to have an approximately linear scaling with $N$, for an optimized purification protocol~\cite{deBone2020}. This result matches the naive expectation of $O(N)$ loss in fidelity going from $N$ Bell pairs to one GHZ state. In the supplement~\cite{supplement}, we work out an example using a specific purification scheme. The impact of noisy operations generally is to limit the maximum achievable fidelity. \newline

\begin{figure}[ht!]
\centering
\includegraphics[width=0.45\textwidth]{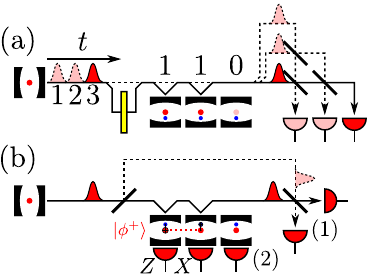}
\caption{Implementation of the hidden matching communication problem. (a) To create the ballot state, the tallyman first generates a single photon in a superposition of time bins. A time-dependent phase is applied to encode the tallyman's randomness. Teleporting to quantum memories, via cavity-assisted photonic gates followed by interferometric measurement, compresses the photonic qudit into a binary qubit code. (b) Similarly, Bell pairs are created by reflecting a time-bin photonic qubit off two cavities, followed by measurement. This entanglement enables teleported CNOT gates, which permute basis states to sample different matchings.}
\label{fig:implementation}
\end{figure}

The most demanding operations that have to be performed are the creation and measurements of the ballot states. As we now discuss, these operations can, nonetheless, be implemented using single quantum emitters strongly coupled to photonic resonators, which function as efficient spin-photon interfaces~\citep{Nguyen2019,Bhaskar2020} (see \autoref{fig:implementation}).   

\emph{Ballot creation.} The key idea to simplify ballot creation involves mapping between unary and binary encodings (see \autoref{fig:implementation}(a)). The ballot state is first encoded in unary using a qudit time-bin encoding where, ideally, a single photon is distributed equally in $n$ time bins. Such a state can either be created using a Raman-type single-emitter scheme~\cite{Lee2018} or approximated using weak coherent pulses. The bit string $x$ is encoded by applying a phase of $\pi$ to photons in select time bins, e.g., by passing them through a phase shifter.

The photonic qudit is subsequently compressed into $\log n$ qubits using spin-dependent reflection from a single-sided cavity~\cite{Reiserer2014,Bhaskar2020}. A resonant photon will be reflected with (without) a $\pi$ phase shift if the emitter qubit is in a state that is uncoupled (coupled) to an excited state by the cavity field~\cite{Duan2004}, which realizes a controlled-phase gate between the qubit and photon. Reflecting the photon off a logarithmic number of such emitter-cavity systems, routing time bins to cavities via the binary representation of the time bin number~\citep{Khabiboulline2019,Khabiboulline2019pra}, will entangle the spin registers with the photonic qudit. Subsequently erasing the time-bin information with an interferometric ($X$-basis) measurement will teleport the ballot state to the registers up to a phase correction determined by the measurement outcome. The operation is heralded on photon detection, which mitigates the detrimental effects of loss in the system to significantly boost the fidelity. Alternatively, the erasing of the time-bin information can be achieved with atomic detection as outlined in Refs.~\citep{Khabiboulline2019,Khabiboulline2019pra}.  

The coherence time of the emitter qubits will limit the size of the ballot state that can be created using the unary to binary encoding outlined above. In particular, the coherence time should be long compared to the duration of the single photon qudit pulse. Furthermore, the complexity of the final interferometric measurement also increases with the size of the qudit state since more time bins need to be interfered. The length of the qudit pulse will be $n\Delta t$ where $n$ is the length of the bit string $x$ and $\Delta t$ is the duration of one time-bin. The latter will be determined by the photon emission time of the emitters and the optical switching rate for routing the time bins to the cavities. For diamond defect centers, Purcell-enhanced photon emission times of about $100$ ps~\citep{Bhaskar2020} would be compatible with time bins on the order of 1 ns, requiring GHz-rate optical switching. For coherence times on the order of 10 ms~\cite{Nguyen2019}, this would give a limit of $n\lesssim10^7$, compatible with thousands of voters. We note that the size of the required quantum memory would only be around $24$ qubits.

Once the ballot state has been encoded in the emitter-cavity registers, it is transferred to a voter via quantum teleportation using the anonymously distributed Bell pairs. For emitters based on color centers in diamond, the Bell pair qubit may be stored in a nearby nuclear spin~\citep{Nguyen2019,Bradley2019}. A Bell measurement is performed using the direct electronic-nuclear spin coupling. The correction bit strings necessary for teleportation can be broadcasted to the voter since they do not contain any information about $x$.

\emph{Ballot measurement.} Voters may sample $\log{n}$ random matchings by measuring one qubit of the ballot state in the $X$ basis and the rest in the $Z$ basis. The other matchings can be obtained by permuting the basis states. The $X$, CNOT, and Toffoli gates are universal in this sense, and the circuit size is bounded by $O(n \log n)$~\citep{Brodsky2004}. In practice, sampling over all matchings may not be necessary, so smaller circuits may suffice, but then the tallyman has more influence on which edge is obtained.

Such operations can be performed with the same type of spin-cavity register as described above. To realize a CNOT gate, Bell states are first established between the spin-cavity systems using spin-photon controlled gates, similar to the ballot state creation except with a photonic time-bin qubit instead of qudit (\autoref{fig:implementation}(b)). Since the ballot state has been transferred to nuclear memory qubits, the probabilistic generation of the Bell states between electronic qubits does not influence the ballot state and can be repeated until successful. Then, applying CNOT gates between the nuclear and electronic spins, measuring the electronic spins, and doing feedback on the nuclear spins results in a teleported CNOT between the nuclear spins~\citep{Chou2018}. Spin rotations on the nuclear spins realize single-qubit gates. \newline

\section{Security Analysis}

\emph{Threat model.}
The voters are partitioned into two sets: honest and dishonest. In our threat model, we trust the set of honest voters to follow the protocol. The same voting protocol is run by multiple tallymen, each having independent randomness, a majority of whom are assumed to be honest. No further assumptions are made and the dishonest tallymen can collude with the set of malicious voters. Additionally, no limits are placed on the computational power of the malicious parties. \newline

We provide definitions for the security properties of a voting protocol, accounting for information-theoretic security. Compared to prior work \cite{Arapinis2021} in the setting of computationally-bounded adversaries, our properties account for the more general setting  of information-theoretic security and therefore provide a stronger security guarantee. We then show that our protocol satisfies these properties; details of the proofs are in the supplement~\cite{supplement}. By $MAJ_N(v_1,\dots,v_N): \{0,1\}^N \to \{0,1\}$ we denote the majority function which takes as input $N$ bits and outputs the bit that occurs most frequently in the input \footnote{Note that ties can be broken arbitrarily.}.  

We begin by formally defining the desirable properties of a voting protocol introduced informally in the beginning of this article.

\begin{defn}[Correctness]\label{defn:correctness}
    A voting protocol $\Pi_{vote}$ is said to satisfy the \emph{correctness} property for some $\epsilon \in (0, 1)$, if given a set of honest voters $\{V_i\}_{i=1}^N$ with votes $\{v_i\}_{i=1}^N$,
    \[
        \Pr[\Pi_{vote}(\{v_i\}_{i=1}^N) = MAJ_N(\{v_i\}_{i=1}^N)] \geq 1 - \epsilon\, .
    \]
\end{defn}
Equivalently, \autoref{defn:correctness} asserts that the protocol outputs the correct answer in the absence of adversarial voters with probability close to unity. \autoref{thm:correctness} demonstrates that our voting protocol satisfies this property.

\begin{defn}[Accountability]\label{defn:accountability}
    A voting protocol $\Pi_{vote}$ is said to satisfy the \emph{accountability} property if for some $\epsilon, \delta > 0$,  
    \[
        \Pr[\text{honest votes are counted}] \ge 1 - \epsilon\, ,
    \]
    and,
    \[
        \Pr[\text{malicious votes are counted}] \leq \delta\, .
    \]
\end{defn}
\autoref{defn:accountability} asserts that the voting protocol always counts votes output by honest voters with high probability and that it does \emph{not} count, with high probability, votes output by any voter that is not honest. \autoref{thm:accountability} demonstrates that our voting protocol satisfies this property.

\begin{defn}[Verifiability]\label{defn:verifiability}
 A voting protocol $\Pi_{vote}$ is said to satisfy the \emph{verifiability} property if every voter $V_i$, $i \in [N]$ can check that their vote $v_i$ is tallied and any auditor can verify the outcome of the election by looking at the transcript of $\Pi_{vote}$.
\end{defn}
\autoref{defn:verifiability} asserts that the protocol is transparent in that it allows voters to check the tallying process. \autoref{thm:verifiability} demonstrates that our voting protocol satisfies this property.

\begin{defn}[Anonymity]\label{defn:anonymity}
    A voting protocol $\Pi_{\text{vote}}$ satisfies the \emph{anonymity} property with parameter $\gamma \in (0, 1)$ if the following holds: Given an honest set of voters $\mathcal{B} \subset \{V_i\}_{i=1}^N $ and a malicious set of voters $\mathcal{A}= \{V_i\}_{i=1}^N\setminus\mathcal{B}$ along with their choice of permutation $\pi: v(\mathcal{B}) \rightarrow v(\mathcal{B})$, such that $\abs{\mathcal{A}} \leq N - 2$, $\forall$ strategies $S$ used by $\mathcal{A}$,
    \begin{equation}
        \Pr_{b\;\overset{R}{\gets}\{0, 1\}}[S(\Pi_{\text{vote}}) = b] \leq \frac{1}{2} + \gamma \,,
    \end{equation}
    where $b$ is a bit that is chosen uniformly at random, such that $\Pi_{vote}$ runs for the configuration of votes chosen by $\mathcal{A}$ for $b = 0$ and the permuted version for $b = 1$. Here, $v(\mathcal{B})$ is the ordered set of votes corresponding to the honest voters.
\end{defn}
\autoref{defn:anonymity} asserts that malicious voters cannot associate a voter's identity to the vote cast by them. More precisely, anonymity is maintained when malicious voters cannot distinguish between any two voting configurations output by honest voters on observing the announcements of the honest voters regardless of the strategy that they use. In practice, imperfect GHZ states reduce the indistinguishability by amount $\gamma$. \autoref{thm:anonymity} demonstrates that our voting protocol satisfies this property.

We now describe how the voting protocol introduced in this article satisfies the requirements of an ideal voting scheme formally introduced above.
First, in the presence of only honest voters, the protocol computes the correct outcome of the voting function.
\begin{thm}[Correctness]\label{thm:correctness}
    The voting protocol described in this article satisfies \autoref{defn:correctness}.
\end{thm}
\begin{sketch}
By construction, assuming that the ballots do not have overlapping edges, the voting protocol outputs the majority winner of the election. The probability that all voters sample the parity of independent bits can be shown to be 
\begin{equation}
    1 - O\left(\frac{N^2}{n}\right)\,,
\end{equation}
from simple combinatorial considerations~\cite{supplement}. To avoid collisions between voters, $n=O(N^2)$, where $n$ is the number of random bits generated by the tallyman, which are encoded in $\log n$ qubits.
\end{sketch}

The second requirement says that no voter can cast more than one vote. Recall that votes are computed as $v=p\oplus a$, where a ballot $((i, j), p)$ is acquired by solving the communication problem and the agreement $a\in \{0,1\}$ is determined by the desired vote. We prove that no voter can output more than one vote that aligns with their intended candidate deterministically, and that extra votes do not bias the election. Casting more than one vote results in all but the first vote being distributed in an unbiased binomial distribution. The influence of the other votes is suppressed by repeating the protocol.
\begin{lem}\label{lem:account}
    Given $N$ copies of the ballot state $\ket{\psi_x}$, a string of $N'\geq N$ votes may be counted with probability
    \begin{equation}
        \Pr[(N'-N)\ \text{ votes}] \leq \frac{1}{2^{N' - N}} \,. \label{eq:guessing}
    \end{equation}
\end{lem}
\begin{sketch}
Votes are only counted if they do not have intersecting vertices. We can therefore consider the task of outputting the overall parity of more than $2N$ bits given access to $N$ ballot states, which bounds the success probability of outputting more than $N$ votes by reduction. We bound the success probability of any measurement strategy correctly distinguishing between states encoding different parities of the random bits. In particular, we show that all extra votes are outputted with $\Pr[v_i=0]=\Pr[v_i=1]=1/2$, i.e., with zero bias~\cite{supplement}.
\end{sketch}

With an appropriate number of repetitions, using the lemma above, the protocol provides a strong degree of accountability.  
\begin{thm}
    [Accountability]\label{thm:accountability}
    The voting protocol described in this article, on repeating $r = O(N, \frac{1}{\delta})$ times with averaging of outputs, satisfies \autoref{defn:accountability}.
\end{thm}
\begin{sketch}
The adversary can technically cast any number of votes. However, outputting extra votes will eventually lead to edge collisions, causing all but the first of their votes to be dropped. As a result, malicious voters can output no more than $O(N)$ vote attempts in a round of the protocol. Repeating $\Pi_{vote}$ using fresh randomness over $O(N)$ rounds and averaging the outcomes suppresses the noise introduced by the extra votes such that they cancel in the computation of the margin~\cite{supplement}. The first vote, conversely, is biased and contributes to the tally.
\end{sketch}

Extending the notion of accountability, the result of the election should be verifiable by both the voters and any external auditor. 
\begin{thm}[Verifiability]\label{thm:verifiability}
    The voting protocol described in this article satisfies \autoref{defn:verifiability}.
\end{thm}
\begin{proof}
    After running $\Pi_{vote}$, a transcript containing the edges $(i_l, j_l)$ along with the agreement bit $a_l$ of every voter is produced. Since $x$ is revealed after every run, a voter can verify that the vote they cast was correctly counted by checking that $(x_{i_l} \oplus x_{j_l}) \oplus a_l = v_l$. An external verifier simply runs the computation in \autoref{thm:accountability} to obtain the margin, by which the election is decided.
\end{proof}

Another critical notion is anonymity: no subset of voters can conspire to identify the votes of the remaining set of voters. Perfect anonymity amounts to only being able to randomly guess the identity of a voter; i.e., the success probability of linking a vote to a voter is $1/N$. In practice, however, anonymity is reduced by a parameter $\gamma$, due to the deviation from perfect fidelity of the entangled states used in the underlying protocols for constructing the queue, distributing the ballot states~\citep{Unnikrishnan2019}, and the collection of votes~\citep{Christandl2005}. This notion of information-theoretic security allows for noisy quantum states~\citep{Unnikrishnan2019}.
\begin{thm}[Anonymity]\label{thm:anonymity}
    The voting protocol described in this article satisfies \autoref{defn:anonymity} with any parameter $\gamma > 0$.
\end{thm}
\begin{sketch}
The voters ensure that they share a high-fidelity GHZ state via the generalized verification protocol. For a GHZ state of given fidelity $\sqrt{1-\gamma^2}$, the deviation from perfect anonymity of the broadcast and summation is shown to be $\gamma$. Similarly, the ballot state is anonymously distributed to the voters up to tolerance $\gamma$~\cite{Unnikrishnan2019}. The total loss of information over the subroutines is bounded, such that the adversary's guess of any voter's vote is close to random~\cite{supplement}:
\begin{equation}
    \Pr\left[\mathcal{A}[S]\text{ guesses }v_k \in \mathcal{B}_0 \right] \le \frac{|\mathcal{B}_0|}{|\mathcal{B}|} + O(\gamma) \,,
\end{equation}
where $\mathcal{B}_0$ is the set of honest voters with $0$-valued votes. By reduction, we show that the probability of identifying voting configurations is similarly bounded.
\end{sketch}

The voting protocol therefore satisfies all the desirable cryptographic properties outlined in Section~\ref{sec:Introduction} and has efficient consumption of communication resources, as summarized in \autoref{thm:main}.
\begin{thm}[Main Theorem] \label{thm:main}
    Given $\epsilon, \delta, \gamma > 0$, voters $\{V_i\}_{i=1}^N$ with votes $\{v_i\}_{i=1}^N$, and the majority voting function $\text{MAJ}_{N}: \{0, 1\}^N \rightarrow \{0, 1\}$, where $\epsilon$ is the tolerance to tag collisions, $\delta$ is the tolerance to failure during repetition, and $\gamma$ is the deviation from perfect anonymity, $\exists n = O(N^2, 1/\epsilon)$ such that,
    \[
        \Pr_{x\;\overset{R}{\gets}\{0, 1\}^n}[\Pi_{vote}^r(x, \{V_i\}_{i=1}^N) = \text{MAJ}_N(\{v_i\}_{i=1}^N)] \geq 1 - \epsilon - \delta\,,
    \]
    where $r = O(N, \frac{1}{\delta})$ is the number of rounds of $\Pi_{vote}$. The scheme satisfies the security properties specified in Definitions \ref{defn:correctness} -- \ref{defn:anonymity} with $\gamma$-anonymity using no more than $O(N^4\log N)$ qubits of communication in $\Pi_{vote}$.
\end{thm}
\begin{sketch}
    The main theorem follows from the proofs of the component theorems and the analysis of the algorithms, included in the supplemental material~\citep{supplement}.
\end{sketch}

\emph{Noisy votes.}  Malicious voters are free to attempt to output as many extra votes as possible. However, there is a limit before losing anonymity. More specifically, in order for the anonymous broadcast to go through, the voters must use the shared Bell pairs. As described earlier, the influence of extra vote attempts is suppressed by repeating the protocol a modest number of times. While a malicious voter can input noise into the broadcast channel, this attack can be detected by the honest voter at the end of the protocol. Our protocol does not guarantee counting \emph{any} vote that a malicious voter may cast. In particular, any voter that attempts to deterministically output more than one vote may have none of their votes counted in the election.\\

\emph{DoS attacks.} Using the fact that a voter does not want to compromise their identity in conjunction with \autoref{thm:correctness}, we can rule out certain families of denial-of-service (DoS) attacks during any phase other than one that uses the anonymous broadcast. Any attempt will either compromise their identity, lead to extra votes that are suppressed during repetitions as shown in \autoref{thm:accountability}, or result in their votes being not counted. As mentioned prior, while noise injected by malicious voters during anonymous broadcast can be detected, the anonymity in the protocol disallows identifying them, leading to a vulnerability to a noise-based DoS attack. This problem is foreseen in Refs.~\cite{Christandl2005,Arapinis2021} and left open for future work. \\

When running $\Pi_{vote}$, any malicious voter $V_i \in \mathcal{A}$, attempting to guess an edge $e = (i', j')$ that intersects with one cast by an honest voter $V_j$, will succeed with vanishing probability (\autoref{thm:correctness}). On the other hand, if $V_i$ observes the broadcasted edge in the tag of an honest voter $V_j$, then $V_i$ can only cast their tag afterwards (as in the order instantiated by the anonymous queue), in which case any vote with overlapping edge is discarded during the execution of $\Pi_{vote}$. If $V_i$ has output edges $\{e'\}$ that do not intersect with those of any other voters and are not obtained from the ballot state, then these votes will be suppressed during repetition (\autoref{thm:accountability}). Lastly, note that if some subset of malicious voters attempts to choose an intersecting set of edges on purpose to compute parities of other edges deterministically, then those extra edges broadcasted in a tag with a vote will be dropped, since they intersect with at least one of the prior edges. In such a case, only the first vote will be counted and subsequent votes will be dropped, which removes incentive for the malicious adversaries to try this attack. These guarantees show that any DoS attack which relies on the protocol aborting due to intersecting edges can be avoided. \newline

\emph{Noise-robustness.} The realistic imperfect fidelity of the Bell states underlying the voting scheme reduces anonymity~\citep{Unnikrishnan2019}, but is testable so the amount of loss is known. Furthermore, our unforgeability result of the quantum ballots remains unchanged for noisy channels. Errors on the ballot state are suppressed over the rounds of the election~\cite[Lemma~S.3]{supplement}. Any final error is detected by the voters at the end of the protocol since they verify whether their vote was counted correctly. While the vote-verification of a group of voters may fail, an election can still be carried out given that this group is small enough compared to the margin of the election.

Operations on the GHZ state should be deterministic and maintain its fidelity, to ensure favorable scaling of the scheme with number of voters. Running the election in parallel by splitting voters in communities of size $O(\log N)$, each having separate tallymen, allows for an error rate that diminishes polynomially in $N$ and, therefore, requires only $\text{poly}(N)$ rounds of repetition (on average) across all communities to have an effectively error-free election. Once a tally is conducted in a community, the tallymen communicate $O(\log N)$ bits to share the margin in their local election. Each tallyman can compute the outcome of the full election locally with this information. This tree structure, combined with network topologies other than a ring, may also help remove adversarial nodes that output DoS attacks. The smaller size of sub-elections also eases the ideal operation of all voters having to participate online at the same time. In addition, voters can in principle submit their votes within some time window, and then the scheme is programmed to run at the end.  \newline

\emph{Example.} To demonstrate the influence of imperfections on the scheme, we analyze a simple example with $N=3$ voters, using current realistic parameters~\cite{Hensen2015,Nguyen2019IntegratedNanophotonic,Pompili2021,Kalb2017,Bradley2019}. Assume voters share Bell pairs with fidelity $F_\mathrm{Bell}$. The fidelity of the resulting GHZ state can be approximated as $F_\mathrm{Bell}^N$. Under a simple, worst-case depolarizing noise model, assume quantum operations and measurements either work perfectly or completely depolarize the state with probability $p$. Then, the fidelity of the GHZ state is approximately $F_\mathrm{GHZ}=F_\mathrm{Bell}^N(1-p)^{f(N)}$, where $f(N)$ is the total number of noisy operations. This fidelity contributes to the anonymity parameter as $\sqrt{1-F_\mathrm{GHZ}^2}$~\cite{Unnikrishnan2019,supplement}. For anonymously casting a vote, where $f(N)=N+1=4,p=0.01,F_\mathrm{Bell}=0.99$, the resulting anonymity leak is $\gamma\approx 0.36$, which is significant compared to $1/2$. Even though the error model is stringent, we see that errors rapidly lead to leaks in anonymity. Nonetheless, given imperfect Bell pairs, the number needed to achieve a target $\gamma$ is polynomial in $N$; see the supplement~\cite{supplement}.

Experimental errors do not diminish the unforgeability of the quantum ballots. However, errors may lead voters to find the wrong parity $p \oplus 1$ with some probability. The influence of this bit flip is suppressed via the $O(N)$ rounds of the protocol by a factor of $\sqrt{N}$, similar to adversarial votes. For example, consider that each gate in the ballot measurement completely depolarizes the state with probability $p$. Then, the averaged bit is flipped with probability $\frac{1}{2}[1-(1-p)^{O(N^2 \log N)}]/\sqrt{N}$. If the vote is flipped at the end of the election, the voter will protest in the verification stage. For $N=3,p=0.01$, and setting constant factors to $1$, the probability evaluates to $0.027$. See the supplement for a more detailed analysis of the bit-flip channel~\cite{supplement}. Thus, errors cause the protocol to abort more often, but the correctness, accountability, and verifiability properties are unaffected.

\section{Conclusion}

In this paper, we present an efficient quantum voting protocol satisfying the desirable security criteria with information-theoretic security (\autoref{thm:main}).  It is a distributed ballot scheme where the solution to a communication complexity problem authenticates voters and encodes their votes. We prove that the quantum encoding prevents forgery of votes, in that only one vote can be recovered from each state. The election outcome can then be computed in a secure multiparty setting using anonymous broadcast. We provide a generalized protocol that enables secure sum, which we use for unconditionally secure collision detection and queuing of voters. An election is run with $O(N^4 \log N)$ qubits of communication, repeated $O(N)$ times to achieve security. Importantly, the $\text{poly}(N)$ scaling is efficient in the number of voters $N$, in contrast to previous schemes. Furthermore, the protocol requires local memories of size $O(\log N)$ qubits for each voter, ensuring that only modestly-sized quantum processors are necessary even for a large number of voters. The reduction in resources, for both communication and computation, removes a fundamental obstacle toward practicality. While beyond the reach of current quantum networks, the amount of quantum processing is modest and many of the key elements of the scheme have already been demonstrated in experiments. Consequently, this work advances the state-of-the-art in cryptography and provides a practical application for quantum networks. The quantum scheme enables capabilities beyond classical voting protocols and may provide hardware-level security.

We can save a factor of $O(N)$ in resources by a more efficient queueing algorithm. In particular, voters adaptively attempt to enter the queue with equal probability, normalized by the amount of remaining spots in the queue. The dimension of the GHZ state is then only $O(N)$. However, a more involved security analysis becomes necessary.

Furthermore, the $O(N)$ overhead introduced by averaging to suppress adversarial votes can potentially be avoided in a tag-based scheme~\citep{supplement}. The tallyman $T$ sends $s = O(\log N)$ copies of the ballot to every voter, and unless the voter can output $s$ parities of the string $x$ at uniformly random locations, the vote is rejected. Then, the probability that an adversary can output votes correctly is bounded by $1/N^c$, for some $c \geq 1$. Communication resources decrease to $\tilde{O}(N^3)$ qubits at the expense of a marginal increase in randomness to $\tilde{O}(N^2)$ bits, suppressing logarithmic factors. However, since a voter makes multiple announcements to cast a single vote, the anonymity of such a protocol becomes challenging to prove in the presence of adaptive adversaries.

Elections in practice may have more than two candidates. More than two choices can be encoded in binary. Then, aggregation would require generalizing the majority vote considered here. We leave extensions of the scheme as a future direction.

Finally, we remark that the voting protocol does not satisfy a property called \emph{Receipt-Freeness}, which asserts that no voter should be able to prove that they voted in a certain way. We believe that ``vote selling'' may still be prevented, and that it is a modest criterion given the voting protocol's other highly desirable guarantees.

\section{Acknowledgments}
We thank Daniel Alabi, Boaz Barak, Mihir Bhaskar, Harry Buhrman, Matthias Christandl, Ignacio Cirac, Kristiaan DeGreve, Peter Love, Carl Miller, Ronald Rivest, Madhu Sudan, Theodor Lukin Yelin, and Leo Zhou for illuminating discussions and useful comments. This work was supported by the NSF, CUA, DOE (grant number DE-SC0020115) and the DARPA ONISQ Program (grant numbers HR001120C0068 and W911NF2010021). E.T.K. was supported by the NSF Graduate Research Fellowship Program (grant numbers DGE1144152 and DGE1745303). J.B. acknowledges funding from the NWO Gravitation Program Quantum Software Consortium.

\bibliography{ms.bib}

\newpage\hbox{}\thispagestyle{headings}\newpage

\onecolumngrid

\renewcommand{\thepage}{S.\arabic{page}}
\setcounter{page}{1}
\renewcommand{\thesection}{S.\Roman{section}}
\setcounter{section}{0}
\renewcommand{\theequation}{S.\arabic{equation}}
\setcounter{equation}{0}
\renewcommand{\thefigure}{S.\arabic{figure}}
\setcounter{figure}{0}
\renewcommand{\thealgocf}{S.\arabic{algocf}}
\setcounter{algocf}{0}
\renewcommand{\thethm}{S.\arabic{thm}}
\setcounter{thm}{0}
\renewcommand{\thelem}{S.\arabic{lem}}
\setcounter{lem}{0}
\renewcommand{\thefact}{S.\arabic{fact}}
\setcounter{fact}{0}
\renewcommand{\thedefn}{S.\arabic{defn}}
\setcounter{defn}{0}

\begin{center} 
\textbf{\large Supplemental Material: Efficient Quantum Voting with Information-Theoretic Security}
\end{center}

\begin{quote}
We begin by providing a summary of important notation and definitions. Then, we specify the algorithms in the quantum voting protocol and provide technical analysis. Furthermore, we complete the proofs of the protocol's cryptographic properties. Finally, we discuss more efficient modifications of the voting scheme.
\end{quote}

\autoref{sec:notations} summarizes common notation that appears throughout the manuscript. In addition, we reproduce the key security definitions.

\autoref{sec:algorithms} elaborates on the algorithms described in the main text. In particular, we state the overarching voting protocol for calculating the election outcome (\autoref{alg:decision}) and the subroutines of anonymous state transfer (\autoref{alg:anon-state-transfer}) and distributed secure sum (\autoref{alg:sum}), which in turn rely on verified GHZ states using \autoref{alg:GHZverification}. We then state the higher-level protocols for anonymously queuing the voters (\autoref{alg:queue}), and conducting one round of voting (\autoref{alg:simple}). Then, we analyze aspects of how the voting protocol runs, bounding the probability of voters sampling intersecting pairs of bits from the quantum ballot state (\autoref{lem:validity}). Finally, we consider noise in the distribution of the quantum ballot state (\autoref{lem:noise}) and bound the probability of a successful run under independent bit flip errors.

\autoref{sec:security} provides the full security proofs for the voting protocol. First, we prove the unforgeability of the quantum ballot, beginning with the result for one copy of the state (\autoref{lem:one-vote}) and afterward in general for multiple copies (\autoref{thm:security}). Then, we bound the influence of adversarial votes when averaging over rounds of the protocol (\autoref{thm:repetition}). Finally, we show anonymity of the voters (\autoref{thm:s.anonymity}).

We introduce variants of the scheme in \autoref{sec:extension}. They enable savings in quantum communication (\autoref{lem:queuing} and \autoref{thm:tags}), but require a more involved security analysis of anonymity.

\section{Notations and Definitions} \label{sec:notations}

Listed below are common terms:
\begin{itemize}
    \item $N$: Number of voters.
    \item $n$: Number of random bits drawn by the tallyman.
    \item $x$: The tallyman's random bit string.
    \item $\ket{\psi_x}=\frac{1}{\sqrt{n}}\sum_{i=0}^{n-1}(-1)^{x_i}\ket{i}$: The ballot state encoded from the random bit string. The label $i$ corresponds to the position of a bit in the string.
    \item $e=(i,j)$: An edge consisting of the labels of two bits as its endpoints.
    \item $p,a,v$: The parity bit $p$, obtained by the XOR of two bits from $x$, is XORed with an agreement bit $a$ chosen by the voter to specify their vote bit $v$.
    \item $O(\cdot)$: Upper bound on asymptotic scaling of a quantity.
    \item $[\cdot]$: Set equivalent to $\{1,2,\ldots,\cdot\}$.
\end{itemize}

We also restate the security definitions.
\begin{defn}[Correctness]
    A voting protocol $\Pi_{vote}$ is said to satisfy the \emph{correctness} property if, for some $\epsilon \in (0, 1)$, given a set of honest voters $\{V_i\}_{i=1}^N$ with votes $\{v_i\}_{i=1}^N$,
    \[
        \Pr[\Pi_{vote}(\{v_i\}_{i=1}^N) = MAJ_N(\{v_i\}_{i=1}^N)] \geq 1 - \epsilon\, .
    \]
\end{defn}
\begin{defn}[{$(\epsilon,\delta)$}-Accountability]
    A voting protocol $\Pi_{vote}$ is said to satisfy the \emph{accountability} property if, for some $ \epsilon, \delta > 0$,  
    \[
        \Pr[\text{honest votes are counted}] \ge 1 - \epsilon\, ,
    \]
    and,
    \[
        \Pr[\text{malicious votes are counted}] \leq \delta\, .
    \]
\end{defn}
\begin{defn}[Verifiability]
 A voting protocol $\Pi_{vote}$ is said to satisfy the \emph{verifiability} property if every voter $V_i$, $i \in [N]$ can check that their vote $v_i$ is tallied and any auditor can verify the outcome of the election by looking at the transcript of $\Pi_{vote}$.
\end{defn}
\begin{defn}[Anonymity] \label{defn:anonymityS}
    A voting protocol $\Pi_{\text{vote}}$ satisfies the \emph{anonymity} property with parameter $\gamma \in (0, 1)$ if the following holds: Given an honest set of voters $\mathcal{B} \subset \{V_i\}_{i=1}^N $ and a malicious set of voters $\mathcal{A}= \{V_i\}_{i=1}^N\setminus\mathcal{B}$ along with their choice of permutation $\pi: \mathcal{B} \rightarrow \mathcal{B}$, such that $\abs{\mathcal{A}} \leq N$, $\forall$ strategies $S$ used by $\mathcal{A}$,
    \begin{equation}
        \Pr_{b\;\overset{R}{\gets}\{0, 1\}}[S(\Pi_{\text{vote}}) = b] \leq \frac{1}{2} + \gamma \,,
    \end{equation}
    where $b$ is a bit that is chosen uniformly at random, such that $\Pi_{vote}$ runs for the configuration of votes chosen by $\mathcal{A}$ for $b = 0$ and the permuted version for $b = 1$.
\end{defn}

\section{Algorithms} \label{sec:algorithms}
We present the pseudocode for all the algorithms along with brief informal explanations for each algorithm.

\begin{algorithm}[h!]\caption{\textsc{Decision}} \label{alg:decision}
    \SetAlgoLined
    \SetKwInput{Input}{Resources}
    \KwIn{$\{v_i\}_{i=1}^N$}
    \KwOut{$\text{MAJ}_N(\{v_i\}_{i=1}^N)$}
    \Input{$O(N^5 \log N)$ Bell pairs}
    \BlankLine
    Initialize $\#0$=$\#1$=0 \;
    \For{$r' \in [r], r=O(N)$}{
    Run \textsc{Voting} (Alg.~\ref{alg:simple}) \;
    \For{$l \in [N]$}{
    \If{$(i_l,j_l) \cap (i_{l'},j_{l'}) = \varnothing$, $\forall l' < l$}
    {\eIf{$v_l=0$}{$\#0=\#0+1$\;}{$\#1=\#1+1$\;}}
    \eIf{$p_l \neq x_{i_l}\oplus x_{j_l}$}
    {$b_l =1$\;}{$b_l =0$\;}
    }
    Compute $B_{r'}=\sum_{i=1}^N b_i \bmod N$ (\textsc{Sum})\;
    }
    Compute averages $\overline{\#0} = \#0/r$, $\overline{\#1} = \#1/r$ and margin $\overline{\#0}-\overline{\#1}$ \;
    Compute average number of complaints $\sum_{r'} B_{r'} / r$ \;
    \If{$\lvert \overline{\#0}-\overline{\#1} \rvert > 2 \sum_{r'} B_{r'} / r$ }{Output winner $(1-\sign(\overline{\#0}-\overline{\#1}))/2$ \;} 
\end{algorithm}

\subsection{Decision Function}

The calculation of the election winner in majority vote relies on averaging over $O(N)$ rounds in order to suppress adversarial votes. The margin of the election is also computed. In each round of the election, voters can anonymously complain if their vote was incorrectly counted. If the number of such complaints is less than half the margin, then the election outcome is unchanged and valid. This process is made explicit in~\autoref{alg:decision}.

\begin{algorithm}[h!]\caption{\textsc{Generalized Verification}}\label{alg:GHZverification}
    \SetAlgoLined
    \SetKwInput{Input}{Resources}
    \KwIn{Random string for every tallyman, $S = |T|\log\left(\frac{4n^2}{\delta\epsilon^2}\right)$, states $\ket{\phi^N_1\cdots \phi^N_{\log(d)}}$}
    \KwOut{\{\texttt{Accept} ($1$),\, \texttt{Reject} ($0$)\}}
    \Input{$O(N^2 \log d)$ Bell pairs}
    \BlankLine
    Mark tallyman $r \in T$ for distribution\;
    $r$ distributes state $\ket{\phi^N_1\dots\phi^N_{\log(d)}}$ distributed to the voters\;
    \For{$t \in \{1,\dots,T\}$}{
        \For{$j \in \{1,\dots,\log(d)\}$}{
            Tallyman $t$ distributes a random $N$-bit string $X_1,\dots,X_N$ which has parity $\sum_{i=1}^N X_i = 0 \bmod 2$ using private channel\;
            Every voter $i \in [n]$ applies, on the $j$-th qubit of their state, a $Z$ operation if $X_i=0$ and a $H$ operation if $X_i=1$\;
            Every voter $i \in [n]$ measures their $j$-th qubit, and sends corresponding outcome $Y_i$ to every tallyman $t' \in T$ via private channel\;
            \If{Tallyman $t$ checks: $\sum_{i=1}^n Y_i \neq \frac{1}{2}\sum_{i=1}^n X_i \bmod 2$}{
                \texttt{Reject}($0$)\;
            }
        }
    Repeat the above $\log\left(\frac{4n^2}{\delta\epsilon^2}\right)$ times with verifying $t'$ chosen independently from the list of tallyman each time\;
    }
    \If{no tallyman rejects}{
        \texttt{Accept}($1$)\;
    }
    \texttt{Reject}($0$)\;
\end{algorithm}

\subsection{Generalized GHZ Verification}

We describe an approach (\autoref{alg:GHZverification}) where we will tensor the GHZ states for qubits $\log(d)$ times to represent a GHZ state for \emph{qudits}. Specifically, we have,
\[
    \ket{\phi^n_0} = \frac{1}{\sqrt{2}}\left(\ket{\phi^k_0}\ket{\phi^{n-k}_0} - \ket{\phi^k_1}\ket{\phi^{n-k}_1}\right)\,,
\]
and so we also have that,
\[
    \ket{\phi^n_0}^{\otimes \log(d)}= \frac{1}{2^{\log(d)/2}}\left(\ket{\phi^k_0}\ket{\phi^{n-k}_0} - \ket{\phi^k_1}\ket{\phi^{n-k}_1}\right)^{\otimes \log(d)}\, .
\]

Note that the above protocol \emph{fixes} a source of randomness $s$ for which \emph{all} $\log(n)$ qubits will be tested simultaneously. This can then be repeated a desirable number of times to bound the success probability using~\autoref{thm:secure-rep-gen} - What is explicitly \emph{not} acceptable is to test with independent sources of randomness $s_1,\dots,s_{\log(d)}$ for each GHZ state that one hopes to verify.

Note that the above modification retains efficiency. To see, observe the following cases:
\vspace{-2mm}
\begin{enumerate}
    \itemsep0.3em
    \item \textbf{Honest case:} If all voters are honest, then given a state with fidelity $\epsilon$ for every $i \in \log(n)$, the probability that every test will be accepted is $(1 - \epsilon^2/4)^{\log(d)} \approx 1 - (\epsilon^2\log(d))/4 + o(1)$. To make this term desirably large, we can run a purification protocol an appropriate number of times~(\autoref{lem:purify-ghz}) and this will allow us to make $(\epsilon^2\log(d))/4 \approx \delta$ for any fixed and sufficiently small $\delta$. This means the protocol will accept with probability $1 - \delta$.
    \item \textbf{Dishonest case:} Note that in this event the probability of accepting decreases by a factor of two to $1 - \epsilon^2/4$. This allows for a gap to emerge and, therefore, the protocol will fail with probability $\epsilon^2/4$ (since this is the probability that \emph{at least} one run fails) which will be $\approx \delta/2$.
\end{enumerate}

\begin{algorithm}\caption{\textsc{Transfer}} \label{alg:anon-state-transfer}
    \SetAlgoLined
    \SetKwInput{Input}{Resources}
    \KwIn{Security parameter S}
    \KwOut{Qubit transmitted from $T$ to $V_{l'}$ with $\epsilon$-anonymity}
    \Input{$O(N^2)$ Bell pairs}
    \BlankLine 
    Initialize $X=0$\;
    \While{$X=0$}{
        $T$ shares an $(N+1)$-qubit GHZ state with voters $\{V_l\}_{l \in [N]}$\;
        \For{$V_l, l \in [N]$}{
            Sample $x \overset{R}{\leftarrow} \{0,1\}^{S}, S=O(\log N)$\;
            Anonymously broadcasts~\citep{Christandl2005} $X=\prod_{i=1}^S x_i$\;
        }
        \eIf{$X = 0$}
        {Random $V_j, j \in [N]$ runs~\autoref{alg:GHZverification}\;
        }
        {$\{V_l\}_{l \in [N]}$ run \textsc{Anonymous entanglement}~\citep{Unnikrishnan2019}, establishing Bell pair $\ket{\phi^+}$ between $V_{l'}, l'\in Q([N])$ and $T$ with $\epsilon$-anonymity\;
        Quantum teleportation from $T$ to $V_{l'}$ using $\ket{\phi^+}$\;}
    }
\end{algorithm}

\begin{algorithm}\caption{\textsc{Sum}} \label{alg:sum}
    \SetAlgoLined
    \SetKwInput{Input}{Resources}
    \KwIn{$\{a_i\}_{i=1}^N,d$}
    \KwOut{$\{a\}_{i=1}^N, a=\sum_{i=1}^N a_i \bmod d$}
    \Input{$O(N^2 \log d)$ Bell pairs}
    \BlankLine
    $T$ shares $\lceil \log d \rceil$ copies of an $(N+1)$-qubit GHZ state, verified with \autoref{alg:GHZverification}, with voters $\{V_l\}_{l \in [N]}$\;
    \For{$V_l, l \in [N]$}{
    Apply $Z^{a_l}$, where $Z=\frac{1}{\sqrt{d}}\sum_{j=0}^{d-1}e^{2 \pi i j /d} \ket{j}\bra{j},a_l\in\{0,\ldots,d-1\}$ \;
    }
    \For{$V_l, l \in [N]$}{
    Apply $U_\text{QFT}$, where $(U_\text{QFT})_{ab}= \frac{1}{\sqrt{d}}(e^{2\pi i / d})^{ab}, \quad 0\leq a,b \leq d-1$ \;
    }
    \For{$V_l, l \in [N]$}{
    Measure $Z_l$, with outcome $k_l$\;
    Broadcast $k_l$\;
    }
    \For{$V_l, l \in [N]$}{
    Compute $-\sum_{i=1}^N k_i \bmod d = \sum_{i=1}^N a_i \bmod d$\;
    }
\end{algorithm}

\begin{algorithm}\caption{\textsc{Queue}} \label{alg:queue}
    \SetAlgoLined
    \SetKwInput{Input}{Resources}
    \KwIn{$N,d=O(N^2)$}
    \KwOut{$Q=\sigma([N]), \sigma \in S_N$}
    \Input{$O(N^4 \log N)$ Bell pairs}
    \BlankLine
    \For{$V_l, l \in [N]$}{
    $V_l$ samples $r_l \overset{R}{\leftarrow} [d]$\;
    }
    \For{$i \in [d]$}{
    \For{$V_l, l \in [N]$}{
    \eIf{$r_l=i$}
    {$b_l=1$\;}
    {$b_l=0$\;}
    Compute $B^N, B=\sum_{i=1}^N b_i \bmod N$ (Alg.~\ref{alg:sum})\;
    \uIf{$B=0$}{$\sigma'(i)=0$\;}
    \uElseIf{$B=1$}{$\sigma'(i)=l$\;}
    \Else{Abort\;}
    }
    }
    $j=1$\;
    \For{$i \in [d]$}{
    \If{$\sigma'(i)>0$}{
    $\sigma(j)=\sigma'(i)$\;
    $j=j+1$\;
    }
    }
\end{algorithm}

\begin{algorithm}\caption{\textsc{Voting}} \label{alg:simple}
    \SetAlgoLined
    \SetKwInput{Input}{Resources}
    \KwIn{$\{v_l\}_{l=1}^N$}
    \KwOut{$\{(i_l,j_l),v_l\}_{l=1}^N$ for each $V_l$}
    \Input{$O(N^4 \log N)$ Bell pairs}
    \BlankLine
    Anonymous queue (Alg.~\ref{alg:queue}) $Q$\;
    $T$ distributes $\ket{\psi_x}^{\otimes N}$, where $x \overset{R}{\leftarrow} \{0, 1\}^n$, anonymously (Alg.~\ref{alg:anon-state-transfer}) to $\{V_l\}_{l \in Q}$\;
    \For{$V_l, l \in Q$}{
    Solve $(i,j),p$ from $x$~\citep{Gavinsky2007}\;
    Set agreement bit $a=p \oplus v$\;
    Anonymous broadcast~\citep{Christandl2005} (Alg.~\ref{alg:sum} with $d=2$) of $(i_l,j_l),a_l$ \;
    }
    $T$ broadcasts $x$\;
    \For{$V_l, l \in [N]$}{
    Compute $v_i \equiv p_i \oplus a_i, i \in [N]$\;
    }
\end{algorithm}

\subsection{Anonymous State Transfer}

The algorithm used for anonymous state transfer (\autoref{alg:anon-state-transfer}) is adapted from \cite{Unnikrishnan2019}. A successful run of the protocol establishes one Bell pair between a voter $V_l$ and tallyman $T$ with $\epsilon$-anonymity (\autoref{lem:tally-voter-anon-state}). Testing of the entanglement limits the success probability to $O(1/N)$. Since a GHZ state can be shared with $N$ Bell pairs, anonymous teleportation of one qubit consumes $O(N^2)$ entangled pairs.
\begin{lem}[\cite{Unnikrishnan2019}] \label{lem:tally-voter-anon-state}
    $\forall \epsilon > 0$, given that the distributed GHZ states have fidelity $F \geq \sqrt{1 - \epsilon^2}$, even in the presence of malicious adversaries $\mathcal{B}$,
    \[
        \Pr[\mathcal{B}\text{ guesses }V_l] \leq \frac{1}{\abs{\mathcal{A}}} + \epsilon\, ,
    \]
    where a voter $V_l$ and tallyman $T$ establish a Bell pair.
\end{lem}

\subsection{Secure Sum}

\autoref{alg:sum} computes the distributed secure sum. It relies on the correlations in GHZ states to ``mask'' the inputs, and outputs their sum (modulo $d$). Below, we provide the calculations that show the correctness of the protocol.

We generalize the one-bit anonymous broadcast protocol~\cite{Christandl2005} to qudits. The initial state is
\begin{eqnarray}
    \ket{\text{GHZ}_d}&=&\frac{1}{\sqrt{d}}[\ket{0}^{\otimes N}+\ldots+\ket{d-1}^{\otimes N}] \\
    &=& \frac{1}{\sqrt{d}}\sum_{j=0}^{d-1}\ket{j}^{\otimes N} \,.
\end{eqnarray}

A voter applies $Z^a$, where $a\in\{0,\ldots,d-1\}$ is the announcement:
\begin{equation}
    Z^a\ket{\text{GHZ}_d}=\frac{1}{\sqrt{d}}\sum_{j=0}^{d-1}e^{2 \pi i j a /d} \ket{j}^{\otimes N} \,. \label{eq:Z_GHZ}
\end{equation}
Note that the action is the same no matter which voters apply $Z$; only the number of applications $a$ matters.

Everyone applies $U_\text{QFT}$, which maps $\ket{j}\to \frac{1}{\sqrt{d}} \sum_{k=0}^{d-1} e^{2\pi i j k/d}  \ket{k}$:
\begin{eqnarray}
    U_\text{QFT}^{\otimes N} Z^a\ket{\text{GHZ}_d} &=& \frac{1}{\sqrt{d}}\sum_{j=0}^{d-1}e^{2 \pi i j a /d} U_\text{QFT}^{\otimes N} \ket{j}^{\otimes N} \\
    &=& \frac{1}{\sqrt{d}}\sum_{j=0}^{d-1}e^{2 \pi i j a /d} (U_\text{QFT} \ket{j})^{\otimes N} \\
    &=& \frac{1}{\sqrt{d}}\sum_{j=0}^{d-1}e^{2 \pi i j a /d} \bigotimes_{l=1}^N(\frac{1}{\sqrt{d}} \sum_{k_l=0}^{d-1} e^{2\pi i j k_l/d}  \ket{k_l}) \\
    &=& \frac{1}{d^{(N+1)/2}}\sum_{j=0}^{d-1}e^{2 \pi i j a /d} \sum_{k_l=0\,\ \forall l}^{d-1} \prod_{l=1}^N e^{2\pi i j k_l/d} \bigotimes_{l=1}^N  \ket{k_l} \\
    &=& \frac{1}{d^{(N+1)/2}}\sum_{j=0}^{d-1}e^{2 \pi i j a /d} \sum_{k_l=0\,\ \forall l}^{d-1} e^{2\pi i j \sum_{l=1}^N k_l/d} \bigotimes_{l=1}^N  \ket{k_l} \\
    &=& \frac{1}{d^{(N+1)/2}} \sum_{j,k_l=0\,\ \forall l}^{d-1} e^{(2\pi i j/d) (a+\sum_{l=1}^N k_l)} \bigotimes_{l=1}^N  \ket{k_l} \\
    &=& \frac{1}{d^{(N-1)/2}} \sum_{k_l=0\,\ \forall l}^{d-1} \delta(a=-\sum_{l=1}^N k_l) \bigotimes_{l=1}^N  \ket{k_l} \,.
\end{eqnarray}

Everyone measures this state in the computational basis. For any measurement outcome, $a=-\sum_{l=1}^N k_l$. Thus, if everyone announces their measurement outcome $k_l$, anyone can sum them together to obtain $a$. Note that each $k_l \in \{0,\ldots,d-1\}$ is uniformly distributed. In the protocol where the input is applied classically ($a=0$), the measurement outcomes are used as a shared key.

\subsection{Anonymous Queuing \& Voting}

\autoref{alg:queue} instantiates the anonymous queue that delegates the order in which the voters cast their vote. Every voter samples randomly a slot in $d=O(N^2)$ positions. Then, the distributed secure sum is used to check that none of the voters collided, aborting otherwise.

\autoref{alg:simple} is the actual protocol used by the voters to cast their votes. First, the tallyman anonymously distributes (\autoref{alg:anon-state-transfer}) the hidden-matching state to every voter. Then, the voters anonymously broadcast \cite{Christandl2005} (\autoref{alg:sum} with $d=2$) their tags, which consist of an agreement bit as well as the indices of the measured edge, in the order instantiated by the anonymous queue (\autoref{alg:queue}). The tallyman finally broadcasts the random string and the votes are computed for every voter.

\subsection{Resource Consumption}

One round of the voting protocol has total quantum communication of $O(N^4\log N)$ qubits, enumerated below.
\begin{itemize}
    \item The distributed secure sum consumes $O(N\log N)$ bits of entanglement, performed $O(N^2)$ times to construct the queue. An extra overhead of $O(N)$ rounds is needed to verify the quality of the GHZ states.
    \item The quantum superposition state compresses the tallyman's $n$ bits of randomness in $\log n$ qubits. To avoid intersecting edges between voters, $n=O(N^2)$ (\autoref{lem:validity}). Copies of the state are communicated to $N$ voters, consuming a GHZ state per qubit~\citep{Christandl2005} (with an overhead of $O(N)$ for verifying the entanglement), which may be distributed with consumption of $N$ Bell pairs~\citep{Komar2014}.
    \item The anonymous broadcast protocol requires a ring of maximally entangled states connecting the voters, consuming $O(N)$ Bell pairs per broadcasted bit. Test measurements verifying the entanglement cost $O(N)$ overhead. A single ballot and vote constitute $2\log n +1$ bits.
\end{itemize}

The number of rounds, used to amplify security by removing the influence of adversarial votes, is $O(N)$ in our scheme, in contrast to comparable protocols that are susceptible to double-voting if the number of rounds is not exponential in the number of voters~\citep{Hillery2006,Bonanome2011,Arapinis2021}.

\subsection{Edge Collision}

The following lemma bounds the number of edges that can share an intersecting vertex between two voters given a ballot state. As a result, for $n=O(N^2)$, adversarial voters can guess $O(N)$ ballot configurations before colliding with another voter. More specifically, \autoref{lem:validity} implies that, for any $\alpha > 0$, if a single voter outputs $O(N^{1 + \alpha})$ votes, the probability of a shared vertex between some pair of voters becomes $O(N^{\alpha})$ which is much greater than $\epsilon$ for sufficiently large $N$, and causes the protocol to almost certainly fail.

\begin{lem}\label{lem:validity}
    Given $n = O(N^2/\epsilon)$, $x \overset{R}{\leftarrow} \{0, 1\}^n$ and two voters $V, V'$ with perfect matchings on $n$ vertices $M \overset{R}{\leftarrow} \mathcal{M}_{\frac{n}{2}}, M' \overset{R}{\leftarrow} \mathcal{M}_{\frac{n}{2}}$ and edges $e \overset{R}{\leftarrow}M, e' \overset{R}{\leftarrow}M'$, the probability that they share a vertex is,
    \begin{equation}
        \Pr[e'\cap e \neq \varnothing \mid e, e'] \leq O\left(\frac{N^2}{n}\right) \leq \epsilon\,.
    \end{equation}
\end{lem}
\begin{proof}
Since each edge $e$ is drawn by a voter \emph{i.i.d.},
\begin{equation}
\Pr[e \leftarrow V, e' \leftarrow V'] = \Pr[e \leftarrow V]\Pr[e' \leftarrow V']\,.
\end{equation}
The probability that two voters sample an edge with intersecting vertices is equivalent to the number of edges with a shared vertex given two matchings $M, M'$ that are sampled independently normalized by the number of matchings. We denote the event of a shared vertex between two voters $V, V'$ that have sampled edges $e, e'$ as $\mathcal{E}$. Since we want to bound it for every pair of voters, we begin by first applying a union bound over the number of distinct voter pairs to the event $\mathcal{E}$,
\begin{eqnarray}
    \Pr[\displaystyle\bigcup_{V \neq V'}\mathcal{E}] &\leq& \displaystyle\sum_{V\neq V'}\Pr[\mathcal{E}] \\
    &=& \displaystyle\sum_{V\neq V'}\Pr[e, e', \mathcal{E}]\\
    &=& \displaystyle\sum_{V\neq V'}\Pr_{e \overset{R}{\leftarrow} M}[e]\Pr[e', \mathcal{E}| e]\\
\end{eqnarray}
    We now use conditional probabilities to split up the joint distribution $\Pr[e, e', \mathcal{E}]$ and note that,
    \[
        \Pr_{e \overset{R}{\leftarrow}\mathcal{M}}[e] = \Pr_{e, M}[e, M] = \sum_{M \in \mathcal{M}_{\frac{n}{2}}}\Pr[e|M]\Pr[M]\, .
    \]
    Using the above, we expand as,
\begin{eqnarray}
    \Pr[\displaystyle\bigcup_{V \neq V'}\mathcal{E}] &\leq& \displaystyle\sum_{V \neq V'}\sum_{M \in \mathcal{M}_{\frac{n}{2}}}\Pr_{e \overset{R}{\leftarrow} M}[e|M]\Pr_{M \overset{R}{\leftarrow} \mathcal{M}_{\frac{n}{2}}}[M]\Pr[e', \mathcal{E}| e] \\
    &=& \displaystyle\sum_{V \neq V'}\sum_{M \in \mathcal{M}_{\frac{n}{2}}}\Pr_{e \overset{R}{\leftarrow} M}[e | M]\Pr_{M \overset{R}{\leftarrow} \mathcal{M}_{\frac{n}{2}}}[M]\sum_{M' \in \mathcal{M}_{\frac{n}{2}}}\Pr_{e' \overset{R}{\leftarrow} M'}[e', \mathcal{E} | e, M']\Pr_{M' \overset{R}{\leftarrow} \mathcal{M}_{\frac{n}{2}}}[M'] \,.
\end{eqnarray}
After applying the union bound over the $\binom{n}{2}$ pairs, marginalizing over the sums of choosing a matching $M$ uniformly at random, and choosing the first edge $e$ uniformly at random from the $\frac{n}{2}$ edges in a matching $M$, we compute the probability that the second edge $e'$ chosen uniformly at random from another random matching $M'$ has an intersecting vertex.
\begin{eqnarray}
   \Pr[\displaystyle\bigcup_{V \neq V'}\mathcal{E}] &\leq& \frac{2\binom{N}{2}}{n}\Pr_{e' \overset{R}{\leftarrow} M'}[e', \mathcal{E}| e, M'] \\
    &\leq&\frac{2\binom{N}{2}}{n}(2n-3)\Pr[e'\in M']\,.
\end{eqnarray}
In the above equation, we note that the number of intersecting edges $e'$ with some fixed $e$ are $2n - 3$, and compute the probability of the intersecting edge as the fraction of matchings over $n-2$ vertices to those over $n$ vertices (fixing some intersecting edge).
\begin{eqnarray}
    \Pr[\displaystyle\bigcup_{V \neq V'}\mathcal{E}] &\leq& 4\binom{N}{2}\frac{n-\frac{3}{2}}{n}\frac{|\mathcal{M}_{\frac{n-2}{2}}|}{|\mathcal{M}_{\frac{n}{2}}|} \\
    &=& 4 \binom{N}{2} \frac{n-\frac{3}{2}}{n}\frac{2(n-2)!(\frac{n}{2})!2^{\frac{n}{2}}}{n!(\frac{n}{2}-1)!2^{\frac{n}{2}}} \\
    &=& 8\binom{N}{2}\frac{n-\frac{3}{2}}{n}\frac{\frac{n}{2}}{n(n-1)} \\
    &\leq& 8\binom{N}{2}\frac{1}{n} \\
    &=& O\left(\frac{N^2}{n}\right) \,,
\end{eqnarray}
where we proceed by explicitly counting the number of matchings in $\mathcal{M}_{\frac{n-2}{2}}$ and $\mathcal{M}_{\frac{n}{2}}$, and setting $n = O(N^2/\epsilon)$ to be the choice for the number of random bits that the tallyman uses to draw $x$.
\end{proof}

\subsection{Bit-Flip Errors}

We demonstrate that the effect of independent bit-flip errors on the ballot state is independent of the amount of random bits $n$ (\autoref{lem:noise}). The expected number of repetitions so that the voters can cast valid votes is $\exp(N)$ with probability $\geq 1 - \exp(-\epsilon'^2n)$. However, note that the expected number of voters with corrupted parities when $N$ ballot states $\ket{\psi_x}$ are distributed is $\leq (\frac{1}{2} + \epsilon')n$. If the margin of victory is greater by some constant amount, then it suffices to have far fewer rounds of repetition since the probability that the same voter has errors in their parity computation also decreases exponentially, and so every round will likely have different voters with errors. By running the election in $O(N/\log N)$ communities of size $O(\log N)$ each, we only incur $1/\text{poly}(N)$ probability of single-qubit Pauli $X$ errors (in every community). Therefore, with $\text{poly}(N)$ repetitions on expectation, an error-free run is conducted. \\

\begin{lem}\label{lem:noise}
    Given $N$ copies of the state $\ket{\psi_x}$ transferred under a bit-flip channel $X_p$ independently for every qubit and some arbitrary $\epsilon' > 0$, the probability that $N$ votes are cast in accordance with their preference is
    \[
        \Pr_p[N\ \text{votes}] \geq \Omega(2^{-N})\,,
    \]
    provided that $x$ is $\epsilon'$-balanced (has roughly equivalent 1s and 0s). The error channel is
    \begin{equation}
        X_p : \rho \to (1-p) \rho + p (X \rho X) \,,
    \end{equation}
    for single-qubit density matrices $\rho$.
\end{lem}
\begin{proof}
    Choose some arbitrarily small $\epsilon' > 0$. We define $\#0(x)$ and $\#1(x)$ as the number of zeroes and ones in a bit string $x$. Notice that a Chernoff bound \cite{mitzenmacher2017probability} implies that
    \begin{align}
    \begin{split}\label{eq:chernoff-x}
        \#1(x) &\in [(\frac{1}{2} - \epsilon')n, (\frac{1}{2} + \epsilon')n] \,, \\
        \#0(x) &\in [(\frac{1}{2} - \epsilon')n, (\frac{1}{2} + \epsilon')n] \,,
    \end{split}
    \end{align}
    with probability $\geq 1 - \exp(-\epsilon'^2 n)$, which is the precise characterization of $x \overset{R}{\leftarrow} \{0,1\}^n$ being $\epsilon'$-balanced.
    
    Fix some edge $(i, j)$ that a voter $V_l$ chooses (with probability $2/n$) and an $\epsilon'$-balanced $x$. Note that, even under the presence of errors, the probability that the measurement is correct is if $\ket{i} \rightarrow \ket{i'}$ and $\ket{j} \rightarrow \ket{j'}$, such that
    \begin{equation}\label{eq:succ-event-1}
        \left((x_{i'} = x_{i}) \wedge (x_{j'} = x_{j})\right) \vee \left((x_{i'} = 1 \oplus x_{i}) \wedge (x_{j'} = 1 \oplus x_{j})\right) = 1 \,.
    \end{equation}
    Note that the probability of the first event in \autoref{eq:succ-event-1} is simply:
    \begin{align}\label{eq:lower-bd-succ-1}
    \begin{split}
        (&\sum_{t=0}^{\log n}\Pr[x_{i'} = x_i]\Pr[d_H(x_{i'}, x_i) = t])(\sum_{t=0}^{\log n}\Pr[x_{j'} = x_j]\Pr[d_H(x_{j'}, x_j) = t]) \\
        \geq&4\left(\frac{1}{2} - \epsilon'\right)^2\left(\frac{1}{2} - \epsilon' - \frac{1}{n}\right)^2 \,,
    \end{split}
    \end{align}
    where $d_H(x,y)$ denotes the Hamming distance between strings $x$ and $y$. We used the fact that the $X_p$ channel acts on $\ket{i}$ and $\ket{j}$ independently and that $x$ is drawn uniformly at random, since we are in the regime of the Chernoff bound set in \autoref{eq:chernoff-x}. By symmetry,
    \begin{align}
    \begin{split}
        (&\sum_{t=0}^{\log n}\Pr[x_{i'} = 1 \oplus x_i]\Pr[d_H(x_{i'}, x_i) = t])(\sum_{t=0}^{\log n}\Pr[x_{j'} = 1 \oplus x_j]\Pr[d_H(x_{j'}, x_j) = t]) \\
        \geq&4\left(\frac{1}{2} - \epsilon'\right)^2\left(\frac{1}{2} - \epsilon' - \frac{1}{n}\right)^2 \,.
    \end{split}
    \end{align}
    
    The total probability that a single voter $V_l$ measures the correct parity of an edge $(i, j)$ chosen uniformly at random from $\mathcal{M}_{\frac{n}{2}}$ given
    \begin{equation}\label{eq:pauli-action}
        \ket{\psi_x} \xrightarrow{X_p^{\otimes \log n}} \ket{\psi'_x}
    \end{equation} 
    is
    \begin{equation}\label{eq:success-one-voter}
        \Pr[p_{ij} = p_{i'j'}] \geq 8\left(\frac{1}{2} - \epsilon'\right)^2\left(\frac{1}{2} - \epsilon' - \frac{1}{n}\right)^2 \,.
    \end{equation}
    Since there are $N$ independent copies of $\ket{\psi_x}$, the probability of valid votes being cast in accordance with the voters' preferences is
    \begin{align}
    \begin{split}
        \Pr[\text{N votes}] = &\geq 2^{3N}\left(\frac{1}{2} - \epsilon'\right)^{2N}\left(\frac{1}{2} - \epsilon' - \frac{1}{n}\right)^{2N} \\
        &= 2^{-N}(1 - 2\epsilon')^{2N}(1 - 2(\epsilon' + \frac{1}{n}))^{2N} \\
        &= 2^{-N}(1 - 4\epsilon' N + o(1))(1 - 4(\epsilon' + \frac{1}{n})N + o(1)) \\
        &= 2^{-N}\cdot \Omega(1) \,,
    \end{split}
    \end{align}
    if we choose $\epsilon' = \frac{1}{cN}$, for some arbitrarily large $c > 0$.
\end{proof}

\section{Security Proofs} \label{sec:security}

Verifiability is proven directly in the main text. Correctness follows from \autoref{lem:validity}. The proof of accountability relies on the unforgeability of the quantum ballot. This result is demonstrated for one copy of the state in \autoref{lem:one-vote}, and then we generalize the proof strategy for multiple copies in \autoref{thm:security}. Then, \autoref{thm:repetition} shows the suppression of adversarial votes via repetition, completing the proof of accountability. The proof of anonymity combines results showing security of the summation (\autoref{thm:anonymousBroadcast}) and verification (\autoref{thm:honest-case-gen} and \autoref{thm:dishonest-case-gen}) protocols, bounding total leakage of information over the combined use of all subroutines in the voting scheme (\autoref{thm:s.anonymity}), and reducing from guessing a voter's vote to satisfying the anonymity definition in terms of identifying permutations (\autoref{thm:anonymityGame}).

\subsection{Unforgeability}

\begin{lem} \label{lem:one-vote}
    Given one copy of $\ket{\psi_x}$, a computationally unbounded adversary \textsf{Adv} can output more than one parity $p$ of an edge $(i,j)$ with probability
    \begin{equation}
        \Pr[\mathsf{Adv} \rightarrow (p, (i, j))] \leq 1/2\,.
    \end{equation}
    
\end{lem}
\begin{proof}

We bound the success probability of outputting the parity of a subset $t$ bits and show that it becomes $1/2$ for $t>2$. That is, the adversary cannot forge a vote from one copy of the state better than a coin flip. A similar approach bounds the quantum communication complexity of the problem~\citep{Shi2015}.

Here, the tallyman has $x\in \{0,1\}^{n}$, which is encoded in $\rho(x)=\ket{\psi_x}\bra{\psi_x}$. The adversary outputs a subset $G$ of $[n]$ of size $t$ and the parity of the bits in that subset. Without loss of generality, the adversary's protocol consists of a measurement $\Pi_G$ to output a subset, followed by a measurement of its parity (\autoref{fact:projection}). $G_+$ (or $G_-$) denote the subsets of $\{0,1\}^{n}$ where that parity is $+$ (or $-$). The probability of distinguishing between two states $\rho_+,\rho_-$ that occur with respective probabilities $p_+,p_-$ is at most, under a positive-operator valued measure (POVM),
\begin{equation} \label{eq:distinguish}
    \frac{1}{2}+\frac{1}{2}\lVert p_+ \rho_+ - p_- \rho_- \rVert_\text{tr} \,,
\end{equation}
where the deviation from $1/2$ is denoted $\epsilon_\text{bias}$. Here, $\lVert \cdot \rVert_\text{tr}$ denotes the trace norm. In our case,
\begin{eqnarray}
    p_+ & = & p_-= \frac{1}{2^{n}} \langle \rho, \Pi_G \rangle \,, \\
    \rho_+ & = & \frac{1}{\langle \rho, \Pi_G \rangle} \sum_{x\in G_+} \sqrt{\Pi_G} \rho(x) \sqrt{\Pi_G} \,, \\ 
    \rho_- & = & \frac{1}{\langle \rho, \Pi_G \rangle}\sum_{x\in G_-} \sqrt{\Pi_G} \rho(x) \sqrt{\Pi_G} \,.
\end{eqnarray}
The bias conditioned on projecting onto $G$ is
\begin{eqnarray}
    \epsilon_\text{bias}\rvert_G &\leq& \frac{1}{2} \lVert \frac{1}{2^{n}} \cdot \sum_{x\in G_+} \sqrt{\Pi_G} \frac{\rho(x)}{\langle \rho, \Pi_G \rangle} \sqrt{\Pi_G} - \frac{1}{2^{n}} \cdot \sum_{x\in G_-} \sqrt{\Pi_G} \frac{\rho(x)}{\langle \rho, \Pi_G \rangle} \sqrt{\Pi_G} \rVert_\text{tr} \\
    &=& \frac{1}{2} \frac{1}{2^{n}} \lVert \frac{1}{\langle \rho, \Pi_G \rangle} \sqrt{\Pi_G} (\sum_{x\in G_+}  \rho(x)  -  \sum_{x\in G_-} \rho(x) ) \sqrt{\Pi_G} \rVert_\text{tr} \,.
\end{eqnarray}
The overall bias is
\begin{equation}
    \epsilon_\text{bias}=\sum_G \langle \rho, \Pi_G \rangle \cdot \epsilon_\text{bias}\rvert_G \,.
\end{equation}

To cancel most of the terms for each $x$, pair elements in $G_+$ and $G_-$ that have the minimum Hamming distance of one. Here, $\delta_{ij}$ is the Kronecker delta function. The index of the flipped bit is $k \in G$.
\begin{eqnarray}
    \sum_{x\in G_+}  \rho(x)  -  \sum_{x\in G_-} \rho(x)  &=& \sum_{x\in G_+}  [\rho(x) - \rho(x \oplus 0^{k-1}\Vert1\Vert0^{n-k})]  \\
    &=&  \sum_{x\in G_+}  [\frac{1}{n}\sum_{i,j=1}^{n}(-1)^{x_i+x_j}\ket{i}\bra{j} - \frac{1}{n}\sum_{i,j=1}^{n}(-1)^{x_i+x_j+\delta_{ik}+\delta_{jk}}\ket{i}\bra{j}] \\
    &=&  \sum_{x\in G_+} \frac{1}{n} [\sum_{i\neq k}(-1)^{x_i+x_k}\ket{i}\bra{k} - \sum_{i\neq k}(-1)^{x_i+x_k+\delta_{ik}+\delta_{kk}}\ket{i}\bra{k} \nonumber \\
    &+& \sum_{j\neq k}(-1)^{x_k+x_j}\ket{k}\bra{j} - \sum_{j\neq k}(-1)^{x_k+x_j+\delta_{kk}+\delta_{jk}}\ket{k}\bra{j}]   \\ 
    &=&  \sum_{x\in G_+} \frac{2}{n} [\sum_{i\neq k}(-1)^{x_i+x_k}\ket{i}\bra{k} + \sum_{j\neq k}(-1)^{x_k+x_j}\ket{k}\bra{j} ] \\
    &=&  \sum_{x\in G_+} \frac{2}{n} \sum_{i\neq k}(-1)^{x_i+x_k}(\ket{i}\bra{k} + \ket{k}\bra{i})  \,.
\end{eqnarray}
Now, invoke cancellations between elements of $G_+$, observing that even and odd-parity substrings of length $t-2$ contribute equally, for $t>2$:
\begin{eqnarray}
    \sum_{x\in G_+}  \rho(x)  -  \sum_{x\in G_-} \rho(x) &=& \frac{2}{n}  \sum_{i\neq k}\sum_{x\in G_+} (-1)^{x_i+x_k}(\ket{i}\bra{k} + \ket{k}\bra{i})  \\
    &=& \frac{2}{n} \cdot
    \begin{cases}
    \sum_{i\neq k} [2^{n-t}] (0)  & \text{if } t=1  \\
    \sum_{i \neq k} [2^{n-t}(-2)] (\ket{i}\bra{k} + \ket{k}\bra{i})  & \text{if } t=2  \\
    \sum_{i\neq k} [2^{n-t}(2\cdot \frac{1}{2}\cdot 2^{t-2}-2\cdot \frac{1}{2}\cdot 2^{t-2})] (\ket{i}\bra{k} + \ket{k}\bra{i})  & \text{if } t>2 
    \end{cases} \\
    &=& 
    \begin{cases}
    -\frac{2^{n}}{n}  \sum_{i\neq k} (\ket{i}\bra{k} + \ket{k}\bra{i})  & \text{if } t=2  \\
    0 & \text{if } t\neq2 
    \end{cases} \,.
\end{eqnarray}

Thus, the optimal strategy is to choose $t=2$ for all $G$ in the POVM. For example, picking $\Pi_G=\sum_{i}^{\lvert G \rvert}\ket{i}\bra{i}$ results in overall bias
\begin{eqnarray}
    \epsilon_\text{bias}&=&\sum_G \langle \rho, \Pi_G \rangle \cdot \epsilon_\text{bias}\rvert_G \\
    &=& \frac{1}{2} \sum_G \langle \rho, \Pi_G \rangle \\
    &=& \frac{1}{2} \,,
\end{eqnarray}
since $\sum_G \Pi_G=\mathbbm{1}$ and $\rho$ is a normalized density matrix. That is, only one vote (parity and edge) can be outputted deterministically. Otherwise, success is due to guessing, since $\epsilon_\text{bias}=0$ for $t\neq 2$:
\begin{equation}
    \Pr[\mathsf{Adv} \rightarrow (p, (i, j))] = \frac{1}{2} + \epsilon_\text{bias} = \frac{1}{2} \,.
\end{equation}

\end{proof}

\begin{fact}[\cite{Shi2015}] \label{fact:projection}
    For an arbitrary POVM $\Pi=\{\Pi_{a,b}\}$ indexed by $a\in \mathcal{A}, b\in \mathcal{B}$, one can construct POVMs
    \begin{equation}
        P=\{P_a\}_{a\in \mathcal{A}} \quad \text{and} \quad Q_a=\{Q_{a,b}\}_{b\in \mathcal{B}}\quad \text{for}\ a\in \mathcal{A} \,,
    \end{equation}
    such that for any state $\rho$ that $\Pi$ acts on, applying $P$, followed by $Q_a$, where $a$ is the outcome of $P$, gives the same output distribution. In particular,
    \begin{equation}
        P_a=\sum_{b\in\mathcal{B}} \Pi_{a,b} \quad \text{and} \quad Q_{a,b}=\sqrt{P_a^+}\Pi_{a,b}\sqrt{P_a^+}\,,
    \end{equation}
    where $P_a^+$ is the Moore-Penrose pseudoinverse of $P_a$.
    
    In our case, $\mathcal{A} = \{G \mid G \subset [n], |G| = t\}$ and $\mathcal{B} = \{+, -\}$.
\end{fact}

\begin{thm} \label{thm:security}
    Given $N$ copies of the ballot state $\ket{\psi_x}$, a computationally unbounded adversary \textsf{Adv} can output a string of $N'\geq N$ parities $p_l$ of edges $(i_l,j_l)$, with probability
    \begin{equation}
        \Pr[\mathsf{Adv} \rightarrow \{p_l, (i_l, j_l)\}_{l=1}^{N'-N}] \leq \frac{1}{2^{N' - N}} \,. \label{eq:s.guessing}
    \end{equation}
\end{thm}
\begin{proof}

We now want to show that given $N$ copies of the state $\ket{\psi_x}$, the adversary cannot output successful forgeries, with probability more than coin flips. To do so, we generalize the calculation of the bias in \autoref{lem:one-vote}. Here, the projection is onto a subset $G$ of $[n]^N$. 
\begin{eqnarray}
    \epsilon_\text{bias}\rvert_G &\leq& \frac{1}{2}  \lVert \frac{1}{2^{n}}\sum_{x\in G_+} \sqrt{\Pi_G} \frac{\rho(x)^{\otimes N}}{\langle \rho^{\otimes N}, \Pi_G \rangle} \sqrt{\Pi_G} - \frac{1}{2^{n}}\sum_{x\in G_-} \sqrt{\Pi_G} \frac{\rho(x)^{\otimes N}}{\langle \rho^{\otimes N}, \Pi_G \rangle} \sqrt{\Pi_G} \rVert_\text{tr} \\
    &=& \frac{1}{2} \frac{1}{2^{n}} \frac{1}{\langle \rho^{\otimes N}, \Pi_G \rangle} \lVert \sqrt{\Pi_G} (\sum_{x\in G_+}  \rho(x)^{\otimes N}  - \sum_{x\in G_-}  \rho(x)^{\otimes N} )\sqrt{\Pi_G} \rVert_\text{tr}
\end{eqnarray}

As in \autoref{lem:one-vote}, cancel terms between the states, where the bit flip is applied to index $k\in G$:
\begin{eqnarray}
    \sum_{x\in G_+}  \rho(x)^{\otimes N}  - \sum_{x\in G_-}  \rho(x)^{\otimes N} &=& \sum_{x\in G_+} [\rho(x)^{\otimes N} - \rho(x \oplus 0^{k-1}\Vert1\Vert0^{n-k})^{\otimes N}] \\
    &=& \sum_{x\in G_+} \frac{2}{n^N} \sum_\alpha\sum_{i_\alpha,j_\alpha\neq k}(-1)^{\sum_\alpha (x_{i_\alpha}+x_{j_\alpha})+x_k}\ket{i_\alpha}\bra{j_\alpha}  \label{eq:cancel_t} \\
    &=&  \frac{1}{n^N} [\sum_{x\in G_+} \sum_\alpha\sum_{i_\alpha,j_\alpha\neq k}(-1)^{\sum_\alpha (x_{i_\alpha}+x_{j_\alpha})+x_k}\ket{i_\alpha}\bra{j_\alpha} \nonumber \\
    &-& \sum_{x\in G_-} \sum_\alpha\sum_{i_\alpha,j_\alpha\neq k}(-1)^{\sum_\alpha (x_{i_\alpha}+x_{j_\alpha})+x_k}\ket{i_\alpha}\bra{j_\alpha}]  \label{eq:flip_back} \\
    &=&  \frac{1}{n^N} [\sum_{x\in G_+} \sum_\beta  \sum_{i_\beta,j_\beta} (-1)^{\sum_{i=1}^n x_i}\ket{i_\beta}\bra{j_\beta} \nonumber \\
    &-& \sum_{x\in G_-} \sum_\beta\sum_{i_\beta,j_\beta}(-1)^{\sum_{i=1}^n x_i}\ket{i_\beta}\bra{j_\beta}]  \label{eq:cancel_rest} \\
    &=&  \sum_{x} \frac{1}{n^N} \sum_\beta  \sum_{i_\beta,j_\beta} \ket{i_\beta}\bra{j_\beta}  \\
    &=& 2^n \frac{1}{n^N} \sum_\beta  \sum_{i_\beta,j_\beta} \ket{i_\beta}\bra{j_\beta}  \label{eq:bias_overlapping} \,.
\end{eqnarray}
In \autoref{eq:cancel_t} is the difference for each $x\in G_+$ flipping $x_k$, due to complete interference of phases: $\ket{i_\alpha}\bra{j_\alpha}$ denote the tensor product of basis states, with an odd number of $k$ substitutions; $\alpha$ indexes all such tuples. Crucially, $c\cdot x_k \mod 2 = x_k$ for any odd $c$. The bit is flipped back in \autoref{eq:flip_back}. Inducting the procedure over all bits $x_i$ yields \autoref{eq:cancel_rest}: $\ket{i_\beta}\bra{j_\beta}$ is composed of a tensor product of an odd number of each basis vector; $\beta$ indexes all such tuples. For clarity,
\begin{eqnarray}
    \ket{i_\alpha}\bra{j_\alpha} \equiv \ket{i_1}\ldots\ket{i_N}\bra{j_1}\ldots\bra{j_N} &\quad \text{s.t.} \quad & \lvert\{i_a=k\}\rvert + \lvert\{j_a=k\}\rvert = c \  \text{for} \  c\vert 2=1 \,, \\
    \ket{i_\beta}\bra{j_\beta} \equiv \ket{i_1}\ldots\ket{i_N}\bra{j_1}\ldots\bra{j_N} &\quad \text{s.t.} \quad& \lvert\{i_b=i\}\rvert + \lvert\{j_b=i\}\rvert = c_i \ \text{for} \ c_i\vert 2 =1 \,, \\
    \forall i_a,j_a,i_b,j_b,i &\in& \{1,\ldots,t\} \,.
\end{eqnarray} 
The constraint dictates which terms appear in \autoref{eq:bias_overlapping}, imposing
\begin{equation}
    \epsilon_\text{bias}\rvert_G=0 \quad \text{if} \quad N<t/2 \label{eq:converse} \,.
\end{equation}

In the voting scheme, a valid vote consists of an edge $(i, j)$ and the parity of its bits $p_{i, j}$. The overall bias is
\begin{equation}
    \epsilon_\text{bias}=\sum_G \langle \rho, \Pi_G \rangle \cdot \epsilon_\text{bias}\rvert_G \leq \frac{1}{2} \,.
\end{equation}
By reduction (\autoref{lem:reduction}), a maximum of $N$ votes can be outputted with unity success probability; outputting any more (i.e., a forgery) can only be done through guessing (\autoref{eq:converse}).

\end{proof}

\begin{lem} \label{lem:reduction}
    Evaluating the parity of $t$ bits reduces to the evaluation of $t/2$ pair-wise parities (assuming $t$ even).
\end{lem}
\begin{proof}
    Exploiting associativity of the parity function,
    \begin{equation}
        \bigoplus_{i=1}^t x_i = \bigoplus_{j=1}^{t/2} (x_j \oplus x_{j+1}) \,,
    \end{equation}
    where $x_i\mapsto x_j$ is any permutation of the bits.
\end{proof}

\subsection{Security Amplification}

\begin{thm} \label{thm:repetition}
     $\forall \delta' > 0$, given $r$ = $O(N)$ repetitions of $\Pi_{\text{vote}}$, $\Pr[N'>N\text{ votes are counted}] \leq \delta'$.
\end{thm}
\begin{proof}
    Generic ballots cast have $N' > N$ votes, where $N$ copies of $\ket{\psi_x}$ are distributed. The combined ballots look like
    \begin{equation}
        \Big\{\Big((i_1, j_1), a_{i_1j_1}\Big),\dots,\Big((i_{N'}j_{N'}), a_{i_{N'}j_{N'}}\Big)\Big\} \subset [n]^2 \times \{0, 1\}\,.
    \end{equation}
    \autoref{thm:security} implies that the $N'-N$ adversarial votes align in the desired direction with probability $1/2$ each. To avoid collisions between voters and proceed with the protocol, $N'-N \leq O(N)$, by \autoref{lem:validity}. The following tail bound applies to the votes.
    \begin{lem}[Additive Chernoff Bound~\cite{mitzenmacher2017probability}] \label{lem:Chernoff}
        For $X=\sum_{i=1}^n X_i$, where $X_i$ are independent, identically distributed Bernoulli random variables, taking the value $1$ with probability $p$ and $0$ otherwise,
        \begin{equation}
            \Pr[ \lvert X - \mathbb{E}[X] \rvert \geq \sqrt{n} \delta ] \leq 2 e^{-2 \delta^2} \,. \label{eq:ChernoffAdditive}
        \end{equation}
    \end{lem}
    
    The influence of the adversarial votes is suppressed by averaging. All votes are counted in one round of the protocol, and the average over $r$ rounds is taken for both the $0$ and $1$ votes (\autoref{alg:decision}). By linearity, consider the adversarial votes separately. Denote by $X$ the total number of adversarial $0$ votes (or $1$, by symmetry, since the adversary cannot bias the output):
    \begin{equation}
        X=\sum_{i=1}^r \sum_{j=1}^{\eta_i} x_{ij}\,, 
    \end{equation}
    where $x_{ij}$ corresponds to round $i$ and vote $j$ in that round, out of $\eta_i$ votes. By \autoref{eq:ChernoffAdditive}, the influence of the adversary is bounded by
    \begin{equation}
        \Pr[ \lvert X - r\eta/2 \rvert \geq \sqrt{r \eta} \delta ] \leq 2 e^{-2 \delta^2} \,,
    \end{equation}
    where $\eta=\max_i \eta_i$.
    Dividing by $r$ to take the average over the rounds yields
    \begin{equation}
        \Pr[ \lvert \bar{X} - \eta/2 \rvert \geq \sqrt{\frac{\eta}{r}} \delta ] \leq 2 e^{-2 \delta^2} \,,
    \end{equation}
    where $\bar{X}=X/r$. To limit the absolute deviation to be less than 1 with high probability, set $r=\eta=O(N)$, since a maximum of $O(N)$ adversarial votes can be inputted into one round. More explicitly, for a deviation of $c$, the number of rounds of averaging is $r = \eta/c^2$. 
    
    To compute the margin of the election, subtract the average number of $0$ votes from the average number of $1$ votes. Since the adversarial $0$ or $1$ votes are symmetric and both represented by $X$, their average values cancel within an additive factor $2c=O(1)$. In majority rule, the election is decided by the sign of the margin. 
    
\end{proof}

\subsection{Anonymity}

We first show that the secure sum with ideal GHZ states is perfectly anonymous (\autoref{sec:PerfectSecureSum}). Generally, the loss of anonymity given high-fidelity generalized GHZ states is controlled (\autoref{sec:ImperfectGHZLeak}). Then, we show that the event of obtaining a generalized GHZ state that is not of high-fidelity is vanishing (\autoref{sec:GHZVerification}). Finally, we combine these results to bound the total loss of anonymity in the voting protocol (\autoref{sec:TotalAnonymityLoss}), in terms of winning the permutation game (\autoref{sec:TranslatingIdentifyingPermutations}), which is the main result (\autoref{thm:anonymityGame}). In addition, we demonstrate that high-fidelity GHZ states can be obtained in a polynomial number of rounds of purification (\autoref{sec:TranslatingInfidelityAnonymity}).

\subsubsection{Secure sum with perfect GHZ states} \label{sec:PerfectSecureSum}

First, we show that the secure sum protocol is secure, given a perfect GHZ state. Similar to the qubit case~\cite{Christandl2005}, the state after application of phases by the voters is the same regardless of who performs the operations (\autoref{eq:Z_GHZ}). The rest of the steps in the protocol are symmetric for honest voters. Thus, in the framework of the anonymity game, permutations of the honest voters are indistinguishable to the adversary. In the case of broadcasting after obtaining shared randomness, the voters' outputs all follow the uniform distribution and are thus indistinguishable, as shown below.

We restrict ourselves to the case where a key was successfully generated and show that the distributions of every single voter under announcements are the same. Let $\mathcal{A}$ represent the subset of malicious voters. The malicious voters $\mathcal{A}$ set a voting configuration $\{v_1,\dots,v_N\}$ for all the voters and choose a permutation over the honest voters $\pi: \mathcal{B} \rightarrow \mathcal{B}$ to permute their votes. For simplicity, assume that the voting function is transitive (\autoref{def:transitive}), i.e., it only depends on the number of $0$ and $1$ votes. This class includes majority vote via a decision on the margin. Also assume a binary alphabet, but note that more than two inputs can still be encoded in a bit string.
    
\begin{defn}[Transitive Voting Function]\label{def:transitive}
    The voting function $f_{vote}: \{0,1\}^N \rightarrow \{0, 1\}$ is said to be \textit{transitive} if
    \begin{equation}
        f_{vote}(v_1,\dots,v_N) = f_{vote}(\pi(v_1,\dots,v_N)), \forall \pi \in S_N \,,
    \end{equation}
    where $v_i \in \{0,1\}, i \in [N]\equiv \{1,\ldots,N\}$ are the votes and $S_N$ is the symmetric group on $N$ letters.
\end{defn}
    
Consider the two cases to be distinguished in the anonymity game.
\begin{enumerate}
    \item \textbf{b = 0:}
        
    The voters vote in the configuration set by $\mathcal{A}$. Voter $V_i$ announces $A_i=k_i + c_i \bmod N$ where $k_i\overset{R}{\leftarrow} \{0,\ldots,N-1\}$ and $c_i\in \{0,\ldots,N-1\}$. By injectivity, the probability distributions are the same:
    \begin{equation}
        P(A_i) = P(k_i) \,.
    \end{equation}  
    The distribution of the announcements of the honest voters is then uniform:
    \begin{equation}\label{eq:announcement-distribution}
        (A_1,\dots,A_{\abs{\mathcal{B}}}) \overset{R}{\leftarrow} \mathcal{U}\{0,N-1\}^{\otimes \abs{\mathcal{B}}} \,.
    \end{equation}
        
    \item \textbf{b = 1:}
        
    The votes are exchanged by the chosen permutation $\pi$. Note that since $f_{vote}$ is transitive, applying $\pi$ on the honest voters does not change the outcome of the election and, therefore, the adversary gains nothing by observing the outcome of the election as $f_{vote}(v_1,\dots,v_N) = f_{vote}(\pi(v_1,\dots,v_N))$.
        
    Note that the permutation $\pi: i \mapsto j$ does not change the distribution of the announcement $A_i=k_i + c_j \bmod N$:
    \begin{equation}
        P(A_i) = P(k_i) \,.
    \end{equation}  
    The announcements are then sampled uniformly at random:
    \begin{equation}\label{eq:announcement-permuted-distribution}
        (A_{\pi(1)},\dots,A_{\pi(\abs{\mathcal{B}})}) \overset{R}{\leftarrow} \mathcal{U}\{0,N-1\}^{\otimes \abs{\mathcal{B}}} \,.
    \end{equation}
\end{enumerate}

The distributions of the announcements $A_i, A_{\pi(i)}$ of every voter $V_i$ are the same for every permutation $\pi$. Given that $\Pr[b=0] = \Pr[b=1]$, there is no bias between the two cases, by \autoref{eq:distinguish}. Therefore, the adversary can do no better than flipping a coin.

\subsubsection{Anonymity leak due to imperfect GHZ states} \label{sec:ImperfectGHZLeak}
Now, we demonstrate that, given a GHZ state with a lower bound on the fidelity, the loss in anonymity is bounded.
\begin{thm} \label{thm:anonymousBroadcast}
    If the voters share a state $\ket{\Psi}$ with fidelity $F'(\ket{\Psi})=\max_U F(U\ket{\Psi},\ket{\text{GHZ}_d})\geq \sqrt{1- \epsilon^2}$, where $U$ is a unitary conducted by the adversary, then the probability of the adversary guessing the identity of the broadcaster, out of $K$ honest parties, is
\begin{equation}
    \Pr[\text{guess}]\leq 1/K+O(\epsilon)\,. \label{eq:imperfectGHZLeak}
\end{equation}

    The probability of guessing a different honest party is
\begin{equation}
    \Pr[\text{guess else}]\geq \frac{1-O(\epsilon)}{K}\,. \label{eq:imperfectGHZLeakElse}
\end{equation}
    
\end{thm}
\begin{proof}

Similar to Ref.~\cite{Unnikrishnan2019}, we show that voters' actions do not change the fidelity of the GHZ state too much, so that the loss of anonymity is bounded. The inputting agent applies $Z$ to their share of $\ket{\text{GHZ}_d}$, then $U_\text{QFT}$. This operation is equivalent to $X U_\text{QFT}$. Consider the states
\begin{equation}
    \ket{\Phi_a^N} = X^{a} U_\text{QFT}^{\otimes N}\ket{\text{GHZ}^N_d} \quad \forall a\in [0,d-1] \,.
\end{equation}
We consider honest adversaries in \autoref{lem:anonymityHonest} and then we allow for malicious behavior in \autoref{lem:anonymityDishonest}. Namely, we find that
\begin{align}
    F(\ket{\Psi_{i'}},\ket{\Psi_j}) &\geq 
        \begin{cases}
            \Big[ 3(1-\epsilon^2)^{1/2}-2+\epsilon^2 \Big]^2 \,,\quad \epsilon \leq \sqrt{\frac{3\sqrt{5}-5}{2}}\approx 0.92 \\
            0 \quad \mathrm{else}
        \end{cases} \\
    &\equiv f(\epsilon) = 1-O(\epsilon^2) \,.
\end{align}

The bound on fidelity implies that the distance between states is similarly small. Note that the distance 
\begin{equation}
    D(\ket{\psi},\ket{\phi})=\sqrt{1-F^2(\ket{\psi},\ket{\phi})} \,,
\end{equation}
for pure states $\ket{\psi},\ket{\phi}$. Then, $D(\ket{\Psi_{i}},\ket{\Psi_j}) \leq \sqrt{1- f^2(\epsilon)}$, where $\ket{\Psi_{i}}$ is state after voter $i$ acts to input. For any POVM element $P$ and states $\rho,\sigma$, $\Tr{P(\rho-\sigma)}\leq D(\rho,\sigma)$~\cite{Wilde2017QuantumInformationTheory2ndEd}. Hence, the probability of an adversary to distinguish between the states, corresponding to different inputting voters, is:
\begin{align}
    \Pr[\text{guess}]&= \frac{1}{K} \sum_{i=1}^K \Tr{\Pi_i \ket{\Psi_{i}}\bra{\Psi_{i}}} \\
    &\leq \frac{1}{K} \sum_{i=1}^K [\Tr{\Pi_i \ket{\Psi_{j}}\bra{\Psi_{j}}}+\sqrt{1- f^2(\epsilon)}] \\
    &= \frac{1}{K}+\sqrt{1- f^2(\epsilon)} \\
    &= \frac{1}{K} + O(\epsilon) \,.
\end{align}

The probability of guessing a different honest party follows from completeness. Besides the broadcasting agent, the other honest parties behave the same way and are indistinguishable. Then,
\begin{align}
    \Pr[\text{guess else}]&\cdot(K-1)+\Pr[\text{guess}]\cdot 1 = 1 \\
    \Pr[\text{guess else}]&=\frac{1}{K-1}(1-\Pr[\text{guess}]) \\
    &\geq \frac{1-1/K-O(\epsilon)}{K-1} \\
    &\geq \frac{1-O(\epsilon)}{K} \,.
\end{align}
\end{proof}

\begin{lem} \label{lem:anonymityHonest}
    Assume that the adversaries behave honestly. If $F(\ket{\Psi},\ket{\text{GHZ}_d^N})= \sqrt{1-\epsilon^2}$, then $F(\ket{\Psi_i},\ket{\Psi_j})\geq 1-\epsilon^2$, where $\ket{\Psi_i}$ is the state after the operations of voter $i$.
\end{lem}
\begin{proof}
    The total state can be expanded as
    \begin{equation}
        \ket{\Psi} = (1-\epsilon^2)^{1/4} \ket{\Phi_0^N} + \sum_{i=1}^{d-1} \epsilon_i \ket{\Phi_i^N} + \sum_{i=d}^{d^n-1} \epsilon_i\ket{\widetilde{\Phi}_i^N}\,,
    \end{equation}
    where $\ket{\widetilde{\Phi}_i^N}\,, \forall i\in [d,d^n-1]$ complete an unspecified, orthogonal basis. 
    
    If voter $j$ applies an input operation $X^a$, the state becomes
    \begin{equation}
        \ket{\Psi_j} = (1-\epsilon^2)^{1/4} \ket{\Phi_a^N} + \sum_{i\in [0,d-1] \setminus a} \epsilon_i \ket{\Phi_{i+a \bmod d}^N} + \sum_{i=d}^{d^n-1} \epsilon_i(j) \ket{\widetilde{\Phi}_i^N}\,.
    \end{equation}
    Define
    \begin{equation}
        \ket{\Phi_a^N}_j^\perp \equiv \sum_{i\in [0,d-1] \setminus a} \epsilon_i \ket{\Phi_{i+a \bmod d}^N} + \sum_{i=d}^{d^n-1} \epsilon_i(j) \ket{\widetilde{\Phi}_i^N} \,.
    \end{equation}
    By normalization of the state,
    \begin{equation}
        \Big\lvert \ket{\Phi_a^N}_j^\perp \Big\rvert = 1-(1-\epsilon^2)^{1/2} \,.
    \end{equation}
    
    Then,
    \begin{align}
        F(\ket{\Psi_{i'}},\ket{\Psi_j}) &= \Big\lvert (1-\epsilon^2)^{1/2} + \sum_{i\in [0,d-1] \setminus a} \lvert \epsilon_i\rvert^2 + \sum_{i=d}^{d^n-1} \epsilon_i(i')\epsilon_i(j) \Big\rvert^2 \\
        &= \Big\lvert (1-\epsilon^2)^{1/2} + \bra{\Phi_a^N}_{i'}^\perp\ket{\Phi_a^N}_j^\perp \Big\rvert^2 \\
        &\geq 
        \begin{cases}
            \Big[ 3(1-\epsilon^2)^{1/2}-2+\epsilon^2 \Big]^2 \,,\quad \epsilon \leq \sqrt{\frac{3\sqrt{5}-5}{2}}\approx 0.92 \\
            0 \quad \mathrm{else}
        \end{cases} \,. \label{eq:lowerBoundHonest}
    \end{align}
    The bound of \autoref{eq:lowerBoundHonest} follows from the Cauchy-Schwarz inequality:
    \begin{equation}
        -[1-(1-\epsilon^2)^{1/2}]^2 \leq \bra{\Phi_a^N}_{i'}^\perp\ket{\Phi_a^N}_j^\perp \leq [1-(1-\epsilon^2)^{1/2}]^2  \,.
    \end{equation}
    
\end{proof}

\begin{lem} \label{lem:anonymityDishonest}
    Assume that the adversaries behave maliciously. If $F'(\ket{\Psi}) \geq \sqrt{1-\epsilon^2}$, then $F(\ket{\Psi_i},\ket{\Psi_j})\geq 1-\epsilon^2$, where $\ket{\Psi_i}$ is the state after the operations of voter $i$.
\end{lem}
\begin{proof}
    Denote $\ket{\Psi'}=U\ket{\Psi}$ as the state that maximizes $F'(\ket{\Psi})$. Expand it as
    \begin{equation}
        \ket{\Psi'}=\sum_{i=0}^{d-1} \ket{\Phi_i^K}\ket{\psi_i} + \ket{\chi} \,, 
    \end{equation}
    i.e., a decomposition into honest voter and adversaries, where $\ket{\chi}$ contains both parts. We can decompose the GHZ state similarly:
    \begin{align}
        \ket{\text{GHZ}^N_d}&= \frac{1}{\sqrt{d}}\sum_{i=0}^{d-1} (Z^i \otimes (Z^i)^\dagger) \ket{\text{GHZ}^K_d}\ket{\text{GHZ}^{N-K}_d}\,, \\
        \ket{\Phi_0^N} &= \frac{1}{\sqrt{d}}\sum_{i=0}^{d-1} \ket{\Phi_i^K}\ket{\Phi_{d-i}^{N-K}}\,.
    \end{align}

    Similar to \autoref{lem:anonymityHonest}, the state after the action of honest voter $j$ is
    \begin{equation}
        \ket{\Psi_j}=\ket{\Phi_a^K}\ket{\psi_0} + \sum_{i\in [0,d-1] \setminus a} \ket{\Phi_i^K}\ket{\psi_{i-a \bmod d}}+\ket{\chi(j)} \,.
    \end{equation}
    Then,
    \begin{equation}
        F(\ket{\Psi_{i'}},\ket{\Psi_j})= \Big\lvert \sum_{i=0}^{d-1} \bra{\psi_i}\ket{\psi_i} + \bra{\chi(i')}\ket{\chi(j)} \Big\rvert^2 \,. \label{eq:fidelityDishonest}
    \end{equation}

    We use the promised bound on fidelity to constrain the two components in \autoref{eq:fidelityDishonest}. First,
    \begin{align}
        \sqrt{1-\epsilon^2} &\leq \Big\lvert \bra{\Phi_0^N}\ket{\Psi'} \Big\rvert^2 \\
        &= \frac{1}{d} \Big \lvert \sum_{i=0}^{d-1}\bra{\Phi_{d-i}^{N-k}}\ket{\psi_i} \Big \rvert^2 \\
        &\leq \frac{1}{d} \Big[\sum_{i=0}^{d-1} \Big \lvert \bra{\Phi_{d-i}^{N-k}}\ket{\psi_i} \Big \rvert \Big]^2 \label{eq:triangleIneq} \\
        &\leq \frac{1}{d} \Big[\sum_{i=0}^{d-1} \Big\lvert \bra{\Phi_{d-i}^{N-k}}\ket{\Phi_{d-i}^{N-k}}\Big\rvert^{1/2} \lvert \bra{\psi_i}\ket{\psi_i} \rvert^{1/2} \Big]^2 \label{eq:CSIneq} \\
        &= \frac{1}{d} \Big[\sum_{i=0}^{d-1} \lvert \bra{\psi_i}\ket{\psi_i} \rvert^{1/2} \Big]^2 \\
        &\leq  \sum_{i=0}^{d-1} \bra{\psi_i}\ket{\psi_i} \,, \label{eq:changeNorms}
    \end{align}
    where \autoref{eq:triangleIneq} follows by the triangle inequality, \autoref{eq:CSIneq} from the Cauchy-Schwarz inequality, and \autoref{eq:changeNorms} from Ref.~\cite{sandhu2024Sumofsquaresgaussianprocesses}. 
    
    Second, since $\ket{\Psi'}$ is normalized,
    \begin{equation}
        \lvert \ket{\chi(j)} \rvert \leq 1- \sqrt{1-\epsilon^2} \,.
    \end{equation}
    By the Cauchy-Schwarz inequality,
    \begin{equation}
        -[1- \sqrt{1-\epsilon^2}]^2 \leq \bra{\chi(i')}\ket{\chi(j)} \leq [1- \sqrt{1-\epsilon^2}]^2
    \end{equation}
    
    Finally, we return to \autoref{eq:fidelityDishonest} and bound it as
    \begin{equation}
        F(\ket{\Psi_{i'}},\ket{\Psi_j}) \geq 
        \begin{cases}
            \Big[ 3(1-\epsilon^2)^{1/2}-2+\epsilon^2 \Big]^2 \,,\quad \epsilon \leq \sqrt{\frac{3\sqrt{5}-5}{2}}\approx 0.92 \\
            0 \quad \mathrm{else}
        \end{cases} \,.
    \end{equation}

\end{proof}

\subsubsection{High-Fidelity agreement with the GHZ state via generalized verification} \label{sec:GHZVerification}
The generalized verification protocol ensures that the states of every voter are close to being true GHZ states.~The two components in the analysis are correctness (which corresponds to the honest case) and soundness (which corresponds to the dishonest case). For the former, we show that the generalized protocol accepts with probability close to $1$ provided all voters are honest and share a state that has high-fidelity w.r.t.~the GHZ state~(\autoref{thm:honest-case-gen}). In the latter case, we show that even when there are ($n-k$) adversarial voters, one can bound the probability that the test passes away from $1$ by a fidelity-dependent factor~(\autoref{thm:dishonest-case-gen}). It then follows by straightforward repetition-based arguments that one can ensure high-fidelity GHZ testing with arbitrarily large probability~(\autoref{thm:secure-rep-gen}).

We now reason first about the fact that, in the \emph{Honest} case, given a state with fidelity $\Psi$, the protocol passes with probability $(1 - \epsilon^2/4)$.

\begin{thm}[Honest Case]\label{thm:honest-case-gen}
    Given $D\bigg(\ket{\Psi}, \ket{\Phi^n_0}^{\otimes \log(n)}\bigg) = \epsilon$, 
    \[
        \Pr[\textsf{V accepts}] \leq 1-\epsilon^2 + \frac{\epsilon^2}{2} \leq 1 - \frac{\epsilon^2}{2}\,,
    \]
    for suitably large $n$ and some $0 < \epsilon' < 1$.
\end{thm}
\begin{proof}
    Up to the introduction of the POVMs $P_n$ and $Q_n$, things stay equivalent to the proof in Ref.~\cite{Pappa2012}. However, for the new test, the state behaves as follows,
    \begin{align*}
        \rho &= (1 - \epsilon^2)\ket{\phi^n_0}^{\otimes \log(n)}\bra{\phi^n_0}^{\otimes \log(n)} + \rho^{\perp} \nonumber\, ,
    \end{align*}
    where,
    \begin{align}
        \rho^\perp = \bigg(\sum_{j_1,\dots,j_{\log(n)} \in [2^{n-1}]}\epsilon_{j_1}\dots\epsilon_{j_{\log(n)}}\ket{\phi^n_{j_1}}\dots\ket{\phi^n_{j_{\log(n)}}} \bigg)\bigg( \sum_{j_1,\dots,j_{\log(n)} \in [2^{n-1}]}\epsilon_{j_1}\dots\epsilon_{j_{\log(n)}}\bra{\phi^n_{j_1}}\dots\bra{\phi^n_{j_{\log(n)}}}\bigg)\, ,
    \end{align}
    with the constraint that,
    \begin{align}
        \sum_{j_1,\dots,j_{\log(n)}\in [2^{n-1}]}\epsilon^2_{j_1}\dots\epsilon^2_{j_{\log(n)}} = \epsilon\, .
    \end{align}
    By definition,
    \begin{align*}
        P_n^{\otimes \log(n)} &= \bigg(\ket{\phi^n_0}\bra{\phi^n_0} + \frac{1}{2}I_{S_n} \bigg)^{\otimes \log(n)} \nonumber \\ &= \bigg(\ket{\phi^n_0}^{\otimes \log(n)}\bra{\phi^n_0}^{\otimes \log(n)} + \sum_{i=1}^{ \log(n)}\frac{1}{2^i}\sum_{S \subseteq [\log(n)], |S| = i}I_{S_n}^{\otimes S}(\ket{\phi^n_0}\bra{\phi^n_0})^{\otimes [\log(n)]\setminus S}\bigg)\, ,
    \end{align*}
    where $I_{S_n}$ is orthogonal to $\ket{\phi^n_0}$ and $\ket{\phi^n_1}$, and we will denote
    \[
        A =  \sum_{i=1}^{ \log(n)}\frac{1}{2^i}\sum_{S \subseteq [\log(n)], |S| = i}I_{S_n}^{\otimes S}(\ket{\phi^n_0}\bra{\phi^n_0})^{\otimes [\log(n)]\setminus S}\, .
    \]
    The final probability of acceptance is,
    \begin{align*}
        p_{\textsf{accept}} &= \Tr\left[\rho P_n^{\otimes \log(n)}\right] \nonumber \\
        &= \Tr\left[\rho\ket{\phi^n_0}^{\otimes \log(n)}\bra{\phi^n_0}^{\otimes \log(n)}\right] + \Tr\left[\rho A\right] \nonumber \\
        &= (1-\epsilon^2) + (1-\epsilon^2)\Tr[\ket{\phi^n_0}^{\otimes \log(n)}\bra{\phi^n_0}^{\otimes \log(n)}A] + \Tr[\rho^\perp\ket{\phi^n_0}^{\otimes \log(n)}\bra{\phi^n_0}^{\otimes \log(n)}]
        + \Tr[\rho^{\perp}(A)] \nonumber \nonumber \\
        &= (1-\epsilon^2) + 0\,+\,0\,+ \Tr[\rho^\perp A] \nonumber \\
        &= (1-\epsilon^2) + \frac{1}{2}\Tr[\rho^\perp] \nonumber \\
        &\leq (1 - \epsilon^2) + \frac{1}{2}\epsilon^2\ = 1-\epsilon^2/2\,.
    \end{align*}
\end{proof}

\begin{thm}[Dishonest Case]\label{thm:dishonest-case-gen}
    Let the joint state of all (honest and dishonest) voters be $\ket{\Psi}$. Provided the minimum fidelity is,
    \[
        \epsilon = \min_U D\bigg(U\ket{\Psi}, \ket{\Phi_n^0}^{\otimes \log(n)}\bigg)\, ,
    \]
    the probability the protocol accepts is $\leq 1 - \epsilon^2/2$.
\end{thm}

\begin{proof}
We now decompose the state $\ket{\Psi}$ into the honest and dishonest components as,
\[
    \ket{\Psi} = \sum_{x \in \{0,1\}^{\log(n)}} \underbrace{\left(\bigotimes_{i=1}^{\log(n)} \ket{\phi^k_{x_i}}_i\right)}_{\ket{\phi^k}_{x_1,\dots,x_{\log(n)}}}\ket{\Psi_x} + \sum_{x \in \{0,1\}^{\log(n)}}\left(\sum_{S \subseteq [\log(n) - 1]}\ket{\phi^k}_{x_S}\left(\bigotimes_{i \in [\log(n)]\setminus S}\ket{\chi^{x_i}_i}\right)\right) + \ket{\chi}\, ,
\]
where the honest parts of $\chi \perp \{\bigotimes_{i=1}^{\log(n)} \ket{\phi^k_{x_i}}\}_{x \in \{0,1\}^{\log(n)}}$ and the honest parts of $\ket{\chi^{x_i}_i}$ are orthogonal to $\{\ket{\phi^k_0}_i, \ket{\phi^k_1}_i\}$ and $\ket{\phi^k}_{x_S} = \prod_{i \in S} \ket{\phi^k_{x_i}}_{i}$.

Note that the dishonest parties can apply a unitary $U \in \mathbb{C}^n$ for any of the $\log(n)$ realizations for each of the tests. The goal now becomes to show that the best strategy that can be employed by the dishonest voters $\mathcal{B}$ is a \emph{product} strategy (that is, entanglement does not help). To do this, one uses the law of total probability followed by the Fretchet bound and reduces the bound to winning \emph{one} of the tests.  
The partial trace of $\ket{\Psi}$ reveals that only the case where $|S| = 1$ is important, which is a statistical mixture of the various sub-states.

By the argument in~\cite[Proof of Theorem 3]{Pappa2012}, the optimal strategy is to win every test optimally, since the repetition is with independent randomness. This implies that this statistical mixture should be over states that maximize the winning probability of a \emph{single} run of the protocol, and so the statistical mixture of $\Tr_{[\log(n)]\setminus\{i\}}[\ket{\Psi}]$ should be over the \emph{same} state every time, implying that the optimal state is a tensor product of the single state in~\cite{Unnikrishnan2019}.

By the law of total probability,
\begin{align*}
    &\Pr[Y_H] = \Pr[Y_{H_1},\dots,Y_{H_{\log(n)}}] \le_{\text{Fretchet bound}} \min_{i \in [\log(n)]} \Pr[Y_{H_i}] \\
    &= \min_{i \in [\log(n)]}\frac{1}{2}\left(\Pr[Y_{H_i} \mid x_i = 0] + \Pr[Y_{H_i} \mid x_i = 1]\right)\, ,
\end{align*}
where $Y_{H_i}$ is the event that the adversary guesses the outcomes of the $i$-th test.

Using the fact that the optimal strategy is to \emph{tensor} the optimal single strategy state~\cite{Pappa2012} and then directly applying the calculation of Ref.~\cite[proof of Theorem 2]{Pappa2012}
\begin{align*}
    \Pr[Y_{H_i} \mid x_i = 0] = \frac{1}{2} + \frac{1}{2}\norm{\Tr_{[\log(n)]\setminus\{i\}}[\ket{\Psi}]}_{\mathsf{tr}} \le \Pr[\text{optimal for single round }i] \le_{\text{\cite[Theorem 2]{Pappa2012}}} 1 - \frac{\epsilon_i^2}{2}\, .
\end{align*}    
By the definition of $\epsilon$ provided in~\cite[Proof of Theorem 2]{Pappa2012}, note that,
\begin{align*}
    \epsilon_i &= \min_{U} D\left((I \otimes \dots \otimes \underbrace{U}_{i} \otimes \dots \otimes I)\ket{\Psi},\ket{\phi^n_0}\right) \\
    &\le \sqrt{1 - \langle(I \otimes \dots \otimes \underbrace{R}_{i} \otimes \dots \otimes I)\ket{\Psi} ,\ket{\phi^n_0} \rangle}\, ,
\end{align*}
where $R_i$ is the rotational unitary that upper bounds the loss of precision provided in~\cite[Proof of Theorem 2]{Pappa2012}, and we have used the fact that the optimal strategy is a product strategy. Since each $\epsilon_i$ bounds the probability of winning a test $i$, it suffices to set $\epsilon = \max_{i \in [\log(n)]} \epsilon_i$. 
\end{proof}

\begin{thm}[Security by Repetition]\label{thm:secure-rep-gen}
    Given a tallyman with $|S| \approx \frac{\log(n)}{\delta^2}$ bits of randomness, on repeating the protocol $S$ times as stated below, the probability that any dishonest voter learns any information about an honest voter is $\leq \delta$.
\end{thm}
\begin{proof}
    This holds by exactly the same proof as in~\cite[Theorem 3]{Pappa2012} with $\Pr[C_\epsilon] = \delta$.
\end{proof}

\subsubsection{Bounding Total Loss of Anonymity} \label{sec:TotalAnonymityLoss}

\begin{thm} \label{thm:s.anonymity}
    Given a single run of the voting protocol $\Pi_{vote}$ with vote cast by honest voter $k \in \mathcal{B}$, the probability that for an honest voter $j \in \mathcal{B}$ the malicious agents $\mathcal{A}$ can correctly identify the vote to the voter $j$, given any strategy $S$ is,
    \[
        \Pr[\mathcal{A}[S]\text{ guess }j] \le \frac{1}{\abs{\mathcal{A}}} + \epsilon\, ,
    \] 
    conditioned on the protocol being run on states that are $\epsilon$-close (in fidelity) to GHZ states.

\end{thm}
\begin{proof}
    We bound the probability of correctly guessing the voter in any slot of the queue. The broadcasting of votes and distribution of ballot states separately lead to expressions like in Ref.~\cite{Centrone2021}. What we need is to obtain an expression for the probability of guessing the position of a voter in the queue, given the three sources of leakage: construction of the queue with secure sum, distribution of ballot states with anonymous transfer, and broadcast of votes with anonymous broadcast. We do so with a Fr\'echet-like bound again:
    \begin{align}
        \forall j \in \mathcal{B},\,\forall \text{strategies}~S&, \\
        &\Pr[\mathcal{A}[S]~\mathrm{guess}~j] \\
        &= \Pr[\mathcal{A}[S]~\mathrm{guess}~j,\,\mathrm{queue}] + \Pr[\mathcal{A}[S]~\mathrm{guess}~j,\,\mathrm{transfer}] + \Pr[\mathcal{A}[S]~\mathrm{guess}~j,\,\mathrm{broadcast}] \\
        &\leq \max_{x \in \{\mathrm{queue},\,\mathrm{transfer},\, \mathrm{broadcast}\}} \Pr[\mathcal{A}[S]~\mathrm{guess}~j\mid x]\, .
    \end{align}
    Set $y$ to be the leakiest part of $\Pi$.
    Then, the total probability of guessing the vote of an honest voter in the most informative pipeline in $\Pi$, by the fact that the voting , is given by,
    \begin{equation}
        \Pr[\mathcal{A}~\mathrm{guess}~\mathcal{B} \mid y] \le \max_{j \in |\mathcal{B}|} \Pr[\mathcal{A}[S]~\mathrm{guess}~j\mid y]\, .
    \end{equation}
        
    We now compute the conditional guessing probabilities by referencing the adversarial strategy $S$ via a generic POVM (without loss of generality), conditioned on the starting state in each $x$ being close to a GHZ state (with high fidelity). Such a bound is provided for each part of the protocol and invoked below. 
    
    \paragraph{Transfer} The anonymity of anonymous transfer directly follows for every honest voter $j \in \mathcal{B}$, \emph{conditioned} on the verification protocol in~\autoref{alg:anon-state-transfer} passing. Namely,
    \[
        \Pr[\mathcal{A}[S]~\mathrm{guess}~j\mid \text{~transfer}] \le \frac{1}{|\mathcal{A}|} + \epsilon\, ,
    \]
    where the analysis follows \emph{symmetrically} for receivers as it does for senders via~\autoref{lem:tally-voter-anon-state}.

    Further, note that every honest voter $V_i$ chooses an edge $e_i$ independently and uniformly at random from a random matching $\mathcal{M}_{\frac{n}{2}}$. Therefore, \emph{every} honest voter's announcement is independent and there is no adaptive strategy that $\mathcal{A}$ can use to predict some information of an upcoming announcement, assuming generalized GHZ verification passes with desired fidelity. This argument also shows where the hidden matching state $\ket{\psi_x}$ assists in the protocol $\Pi_{vote}$, as it allows the honest voters $V_i$ the ability to cast their vote with a tag uncorrelated to that of other voters or one that is chosen by the tallyman. Lastly, even with complete knowledge of $x$, the tallyman $T$ cannot distinguish one announcement from the other since the tallyman does not know which edge corresponds to which voter (with guessing advantage more than $\gamma$).

    \paragraph{Broadcast} The loss of anonymity of a voter during the broadcast part of the protocol is bounded by~\autoref{lem:anonymityDishonest}. The bound is identical to that of~\autoref{lem:tally-voter-anon-state} in quantitative strength.
    
    \paragraph{Queueing} The bound on identifying a voter's position in the queue is the same as the bound on the probability of identifying a voter in the broadcast, since the voter only inputs their location once. If there is a collision, the queuing protocol is rerun with fresh randomness.  
\end{proof}

\subsubsection{Translation from Identifying Voters to Guessing Permutations} \label{sec:TranslatingIdentifyingPermutations}

\begin{thm}[Satisfying the anonymity game] \label{thm:anonymityGame}
    The voting protocol $\Pi_{\text{vote}}$ outlined above satisfies the anonymity requirement specified in~\autoref{defn:anonymityS}. \\
    Equivalently, there exists no strategy $\mathcal{S}$ that can be adopted by the dishonest voters $\mathcal{A}$ which can distinguish between two runs of the protocol where the honest votes are permuted under a permutation $\pi$ chosen by $\mathcal{S}$ with probability greater than $\frac{1}{2} + \gamma$. 
\end{thm}
\begin{proof}
    We bound over all possible strategies of the adversary. Consider the (possibly) maximally informative runs to distinguish, where the adversaries choose a strategy $\mathcal{S}$ based on $\pi \in S_{|\mathcal{A}|}$ that flips \emph{every} honest voter's vote. More formally,
    \begin{equation}
        v = (v_1,v_2,\ldots,v_k)\text{ s.t. } \left|\{i\,\mid\, i \in [k], v_i = 1\}\right| = \left|\{i\,\mid\, i \in [k], v_i = 0\}\right|  \,,
    \end{equation}
    and,
    \begin{equation}
        \pi(v_i) \neq v_i\, ,\, \forall i \in [k]\,.
    \end{equation}
    We now follow the procedure in~\cite[Pg. 5 \& 6]{Centrone2021} to bin the votes that are $0$ and $1$ and argue by the law of conditional probability. Note that in order to win the anonymity game, the adversaries \emph{must} be able to associate (in a fixed round of $\Pi_{vote}$) an outputted vote index with a voter $j \in \mathcal{A}$ with probabilities that differ by \emph{at least} $\gamma$, depending on whether $b = 0$ (and the sequence of votes is $v$) or $b=1$ (and the sequence of votes is $\pi(v)$). This is \emph{at least} as hard as distinguishing between whether an honest voter (associated with some index) voted in a set of rounds that outputted $0$ votes and $1$ votes. We denote by $H_0$ the set of honest voters that vote $0$. Likewise, $H_1$ denotes the set of honest voters that vote $1$. Letting $\mathcal{F}$ denote the event that the fidelity of the states in the election is sufficiently large,
    \begin{align}
        &\Pr\left[\mathcal{A}[S]\text{ guesses }v_k \in H_0\,\mid\, \mathcal{F}\right] = \Pr\left[\mathcal{A}[S]\text{ guesses }v_k \in H_0\,\mid\, \mathcal{F},\text{ collision}\right]\Pr[\text{collision}] + \\
        &\Pr\left[\mathcal{A}[S]\text{ guesses }v_k \in H_0\,\mid\, \mathcal{F}, \text{ no-collision}\right]\Pr[\text{no-collision}] \\
        &= \mathsf{neg}(n)\Pr[\text{collision}] + \Pr\left[\mathcal{A}[S]\text{ guesses }v_k \in H_0\,\mid\, \mathcal{F}, \text{ no-collision}\right]\Pr[\text{no-collision}] \\
        &\le \mathsf{neg}(n) + \Pr\left[\mathcal{A}[S]\text{ guesses }v_k \in H_0\,\mid\,\mathcal{F},\text{ no-collision}\right]\,.
    \end{align}
    The probability above may be bounded by using an argument akin to~\cite[Pg. 5 \& 6]{Centrone2021}. Therefore, invoking~\autoref{thm:s.anonymity} immediately implies that,
    \[
        \Pr\left[\mathcal{A}[S]\text{ guesses }v_k \in H_0\,\mid\,\mathcal{F},\text{ no-collision}\right] = \frac{1}{|\mathcal{A}|} + \epsilon + \left(|\mathcal{A}_0| - 1\right)\left(\frac{1}{|\mathcal{A}|} + \epsilon\right)\,,
    \]
    where $\mathcal{A}_0$ denotes the number of honest voters with vote $0$.
    A little bit of algebra immediately simplifies the above to,
    \[
        \Pr\left[\mathcal{A}[S]\text{ guesses }v_k \in H_0\,\mid\,\mathcal{F},\text{ no-collision}\right] \le \frac{|\mathcal{A}_0|}{|\mathcal{A}|} + 2\epsilon\,.     
    \]
    A similar argument can be made to apply to bound the probability that $\mathcal{A}[S]$ can guess some $v_k \in H_1$, conditioned on $\mathcal{F}$ and no collision. Namely,
    \[
        \Pr\left[\mathcal{A}[S]\text{ guesses }v_k \in H_1\,\mid\,\mathcal{F},\text{ no-collision}\right] \le \frac{|\mathcal{A}_1|}{|\mathcal{A}|} + 2\epsilon\,.
    \]
    Consequently,
    \[
        d_{\mathsf{TVD}}\left(\mathbb{P}_{\mathcal{A}[S]},\mathsf{Emp}\left(|\mathcal{A}_0|, |\mathcal{A}_1|\right)\right) \le 4\epsilon\, ,
    \]
    where $d_{\mathsf{TVD}}$ gives the total variation distance, and $\mathsf{Emp}\left(|\mathcal{A}_0|,|\mathcal{A}_1|\right)$ denotes the empirical probability distribution of votes. \\
    Note that, irrespective of whether $b = 0$ or $b = 1$ in the permutation game, the distance of the distribution induced by the adversaries strategy $S$ from the empirical distribution will stay the same. Furthermore,
    \[
        \mathsf{Emp}\left(|\pi\left(\mathcal{A}_0\right)|,|\pi\left(\mathcal{A}_1\right)|\right) \overset{d}{=} \mathsf{Emp}\left(|\mathcal{A}_0|,|\mathcal{A}_1|\right)\, ,
    \] 
    for any $\pi \in S_{|\mathcal{A}_0|}$. Since $d_{\mathsf{TVD}}$ forms a metric and $\Pr[b=0] = \Pr[b=1] = 1/2$, it immediately follows by an application of the triangle inequality that,
    \[
        \Pr_{b \leftarrow \{0,1\}}\left[S(\Pi_{vote})\right] \le \frac{1}{2} + 8\epsilon\,.
    \]
\end{proof}

\subsubsection{Translation from Bell State Infidelity to Anonymity Loss} \label{sec:TranslatingInfidelityAnonymity}

In the following, we analyze an example purification protocol improving the fidelity of entangled states, to find the initial number of noisy Bell states needed to achieve a desired level of anonymity.

\begin{lem}[Efficient purification of GHZ states]\label{lem:purify-ghz}
    Given a sufficient number of Bell-States with fidelity $\geq 1 - \epsilon$, $\log(\mathsf{poly(}n, 1/\epsilon, 1/\gamma))$ runs of the \emph{BBPSSW} protocol will yield GHZ states with fidelity $\geq \sqrt{1 - \gamma^2}$.
\end{lem}
\begin{proof}
    Let the minimum starting fidelity of a Bell state be
    \begin{equation}\label{eq:min-fidelity}
        F^0_{min} = 1 - \epsilon\, ,    
    \end{equation}
    for some $\epsilon > 0$. With this assumption, we can apply the \emph{BBPSSW} purification protocol~\cite{dur2007entanglement} for $r$ rounds. This will increase the fidelity to,
    \begin{equation}\label{eq:fidelity-increase}
        F^r_{min} = \frac{1}{p_{suc}}\cdot\left((F^{r-1}_{min})^2 + \left(\frac{1 - F^{r-1}_{min}}{3}\right)^2\right)\, ,
    \end{equation}
    where $F^{r}_{min}$ is the minimum fidelity across all Bell states after $r$ rounds of purification, and,
    \begin{equation}\label{eq:success-prob}
        p_{suc} = (F_{min}^{r-1})^2 + \frac{2}{3}F_{min}^{r-1}(1 - F_{min}^{r-1}) + \frac{5}{9}(1 - F_{min}^{r-1})^2\, ,
    \end{equation}
    where $p_{suc}$ denotes the probability that the purification succeeds. By a Taylor expansion of \autoref{eq:fidelity-increase} using \autoref{eq:min-fidelity}, we obtain that
    \begin{equation}\label{eq:fidelity-increase-approx}
        F^1_{min} = 1 - \frac{2}{3}\epsilon - \frac{2}{3}\epsilon^2 + O(\epsilon^3) \approx 1 - \frac{2}{3}\epsilon = \frac{1}{3}(1 + 2F^0_{min})\, . 
    \end{equation}
    After some algebra, this yields that,
    \begin{equation}\label{eq:fid-increase-r}
        F^r_{min} = \frac{1}{3^r}(c_r + 2^r\cdot F^0_{min})\, ,
    \end{equation}
    where $c_1 = 1$ and,
    \begin{equation}\label{eq:increase-c}
        c_r = 3^{r-1} + 2\cdot c_{r-1} = 3^r - 2^r\, .
    \end{equation}
    This allows us to further simplify \autoref{eq:fid-increase-r} to its final form as,
    \begin{equation}\label{eq:fid-final-r-increase}
        F^r_{min} = \frac{1}{3^r}(3^r - 2^r(1 - F^0_{min})) = 1 - \left(\frac{2}{3}\right)^r\epsilon\, .
    \end{equation}
     \autoref{alg:simple} consumes $O(N^3\log N)$ GHZ states in total. Additionally, $\Pi_{vote}$ itself is repeated $O(N)$ times, bringing the total number of GHZ states that need to be purified to $O(N^4\log N)$. Note that a GHZ state can be created from $N$ Bell states with no more than $O(N)$ loss of fidelity, assuming perfect gates in the circuit. Such a circuit implementation has been proposed~\cite{Komar2016} for the aforementioned GHZ state preparation. Therefore, by a union bound on \autoref{lem:tally-voter-anon-state}, we see that our total loss of anonymity is,
    \[
        \text{anonymity leak} \leq \epsilon'\cdot O(N^5\log N)\, ,
    \]
    where $\epsilon'$ is the desired fidelity to have $\gamma$-anonymity. To ensure this, the anonymity leak is upper bounded by $\gamma$,
    \begin{equation}\label{eq:anonymity-leak}
        \text{anonymity leak} \leq \epsilon'\cdot O(N^5\log N) = \gamma\, .
    \end{equation}
    Using \autoref{eq:fid-final-r-increase} and the fact that the desired final fidelity by \autoref{eq:anonymity-leak} is,
    \[
        1 - \left(\frac{2}{3}\right)^r\epsilon = F^r_{min} = \sqrt{1 - \frac{\gamma^2}{N^{10}\log^2 N}}\, ,
    \]
    we have the following bound on the number of rounds $r$ required in the purification process,
    \begin{equation}\label{eq:num-rounds}
        r = \log(\frac{4\epsilon}{3\gamma^2}\cdot N^{10}\log^2 N)\, . 
    \end{equation}
Now, because the protocol is a recurrence protocol, the number of expected Bell states required to purify one GHZ state to the desired degree of fidelity is $O(2^r)$. This yields,
\begin{equation}
    \text{\# Bell states} = O\left(2^{\log(\frac{4\epsilon}{3\gamma^2}\cdot N^{10}\log^2 N)}\right) = O\left(\frac{4\epsilon}{3\gamma^2}N^{10}\log^2 N\right) \, .   
\end{equation}
This can be repeated for all $O(N^4\log N)$ GHZ states that will be used in the repetitions of $\Pi_{vote}$.
\end{proof}

\section{Extensions} \label{sec:extension}

We sketch several modifications to the protocol that yield a factor of $\tilde{O}(N)$ reduction in resources each. However, they require more complicated anonymity proofs, which we do not work out.

\subsection{Queue}

We show that the adaptive strategy in the anonymous queuing algorithm requires $O(N)$ rounds, resulting in the consumption of $O(N^3 \log N)$ Bell pairs overall, including verification.

\begin{lem} \label{lem:queuing}
    \begin{equation}
        \Pr[\text{\textsc{Queue} terminates in $O(N)$ rounds}] = 1-1/c \,,
    \end{equation}
     for any constant $c$.
\end{lem}
\begin{proof}
    Let $p_i$ be the probability of successfully filling the $i$th spot in the queue. To succeed, out of the $N-i+1$ unqueued voters, one voter attempts to enter the queue with probability $1/(N-i+1)$ while the rest choose not to. Then, combinatorially,
    \begin{eqnarray}
        p_i &=& (N-i+1)(\frac{1}{N-i+1})^1(1-\frac{1}{N-i+1})^{(N-i+1)-1} \\
        &=& (1-\frac{1}{N-i+1})^{N-i} \\
        &\xrightarrow{N\to \infty} & 1/e \,.
    \end{eqnarray}
    
    The total time $T=t_1 + \ldots + t_N$ is the sum of the times needed to fill the individual spots. Having a geometric distribution, the expected value is $\mathbb{E}(t_i) = 1 / p_i$. By linearity,
    \begin{eqnarray}
        \mathbb{E}(T) &=& \sum_{i=1}^N \mathbb{E}(t_i) \\
        &=& \sum_{i=1}^N 1/p_i \\
        &=& \sum_{i=1}^{N-1} (\frac{N-i+1}{N-i})^{N-i} + 1 \\
        &\xrightarrow{N\to \infty} & N \cdot e \,.
    \end{eqnarray}
    
    Markov's inequality~\cite{mitzenmacher2017probability} bounds the probability of a long run time, for any $c>0$:
    \begin{equation}
        \Pr[T\geq c e N] \leq 1/c \,.
    \end{equation}
\end{proof}

\subsection{Tag}

Here, we describe a modification of our protocol that saves a factor of $\tilde{O}(N)$ in the total resource count.

\begin{figure}[ht]
\centering
\includegraphics[width=0.45\textwidth]{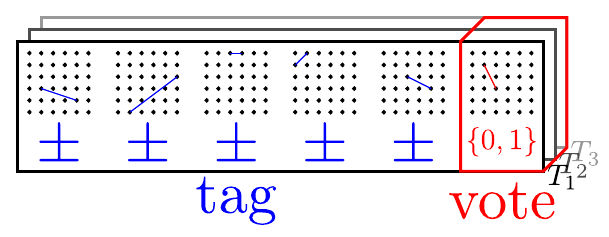}
\caption{To amplify security, voters receive several copies of the ballot state from multiple tallymen. A vote is valid if the additional parities are correct. The outcomes from multiple tallymen are compared to suppress cheating.}
\label{fig:tag}
\end{figure}

Each of $t$ tallymen may send $s$ extra copies of the ballot state to each voter, introducing a tag consisting of random pairs of bits and their parities (Fig.~\ref{fig:tag}):
\begin{equation}
    \Big(((i_1, j_1), p_{i_1j_1}),\dots,((i_{s}j_{s}), p_{i_{s}j_{s}}) \Big) \subset [n]^2 \times \{0, 1\}\,.
\end{equation}
For the vote to be counted, the tag has to be correct. Extending Eq.~\ref{eq:s.guessing}, the probability to pass authentication without access to the ballot states is exponentially small in the length of the tag:
\begin{equation}
    \Pr[(N'-N)\ \text{votes}] \leq \frac{1}{2^{st}}\frac{1}{2^{N' - N}} \,. \label{eq:tag}
\end{equation}
This feature prevents attacks where unverified adversaries attempt to force the protocol to abort. With tags from multiple tallymen, it forces them to cooperate to verify votes. The resource consumption becomes $O(st N \log (sN))$, times a factor of $O((stN)^2)$ for anonymous transmission. For security, the number of copies in the tag is $s=O(\log N)$ (\autoref{thm:tags}). The number of tallymen $t$ can be some constant greater than one to compare outcomes.

\begin{thm} \label{thm:tags}
     $\forall \delta' > 0$, given $s$ = $O(\log N)$ copies of $\ket{\psi_x}$, $\Pr[N'>N\text{ votes are counted}] \leq \delta'$.
\end{thm}
\begin{proof}
    Given $s$ additional copies of the ballot state per voting ballot, the probability of casting adversarial votes is suppressed by a factor $1/2^s$. Extending the reasoning of \autoref{thm:repetition}, each voter has $O(sN)$ such attempts, by \autoref{lem:validity}. To limit the influence on the election to one vote across all voters,
    \begin{equation}
        \frac{sN}{2^s}\leq \frac{1}{N} \,.
    \end{equation}
    Then, $2^s/s \geq N^2$, so asymptotically, a tag of length $s=O(\log N)$ suffices.
\end{proof}    

\end{document}